\documentstyle[aps]{revtex}

\begin{document}
\title{Spherically Symmetric Non Linear Structures}
\author{Esteban A. Calzetta\thanks{%
email: calzetta@iafe.uba.ar} and Alejandra Kandus\thanks{%
email: kandus@iafe.uba.ar}}
\address{Instituto de Astronom\'\i a y F\'\i sica del Espacio (IAFE), c.c. 67, suc\\
28, (1428) Buenos Aires, Argentina and Departamento de F\'\i sica, Facultad\\
de Ciencias Exactas y Naturales, UBA, Ciudad Universitaria, (1428) Buenos\\
Aires, Argentina.}
\maketitle

\begin{abstract}
We present an analytical method to extract observational predictions about
non linear evolution of perturbations in a Tolman Universe. We assume no a
priori profile for them. We solve perturbatively a Hamilton - Jacobi
equation for a timelike geodesic and obtain the null one as a limiting case
in two situations: for an observer located in the center of symmetry and for
a non-centered one. In the first case we find expressions to evaluate the
density contrast and the number count and luminosity distance vs redshift
relationships up to second order in the perturbations. In the second
situation we calculate the CMBR anisotropies at large angular scales
produced by the density contrast and by the asymmetry of the observer's
location, up to first order in the perturbations. We develop our argument in
such a way that the formulae are valid for any shape of the primordial
spectrum.
\end{abstract}

\section{Introduction}

The study of large - scale structure formation represents one of the most
exciting research fields in Cosmology. The currently accepted view is that
the structures we observe today trace their origin to primeval density
inhomogeneities, generated by zero point fluctuations in the scalar field
responsible for Inflation. The evolution of these inhomogeneities, from
certain initial values ($<<1$) to their present distribution in stars,
galaxies, clusters of galaxies and so on, involves complex non linear
hydrodynamic and gravitational processes. Due to this complexity, it has not
been possible to study this evolution from a completely analytical
perspective \cite{Zeldovich},\cite{Roman},\cite{Shandarin}.

Given the impossibility at present (and possibly in principle \cite{Hobill})
to find a general exact analytical solution to Einstein equations, our
understanding of processes such as the clustering of galaxies is largely
dependent on numerical techniques. Of the known exact solutions to those
equations, most of them presuppose the presence of certain symmetries of
space-time and/or energy density \cite{MacCallum}, \cite{Krasinski}. One of
them is the Tolman solution, which describes an inhomogeneous Universe
filled with pressureless matter and spherically symmetric around a point.
This solution is completely characterized by two time independent functions $%
f^2(r)$ and $F(r)$. This last one can be interpreted as the mass contained
in a sphere and $f^2(r)$ as the mechanical energy of a shell, both of radius 
$r$.

In spite of its limitations, the Tolman solution has been fruitfully used to
study a great variety of effects related to the formation of large scale
structures: formation of large scale voids, anisotropies in the microwave
background radiation, possible fractal distribution of galaxies, etc. \cite
{voids}, \cite{Paczynski}, \cite{Panek}, \cite{Ribeiro}, \cite
{Moffat-Tatarski}, \cite{Fullana}\cite{langlois}. But all (or almost all) of
these studies are based on the numerical integration of the corresponding
equations (mainly the equation for null geodesics), and often simplified
expressions for $f^2(r)$ and $F(r)$ are employed, sometimes with no other
reason than to facilitate the numerical calculations.

The goal of this paper is to present an analytical method to study the non
linear evolution of density perturbations in a matter dominated Universe,
viewed as an instance of the Tolman Universe, laying the emphasis in a
physically motivated choice for the Tolman functions. Concretely, we shall
assume the Tolman functions to be such that at early times, in the linear
regime, the model be equivalent to a spatially flat Friedmann - Robertson -
Walker (FRW) matter dominated Universe with growing perturbations.
Otherwise, we shall keep the Tolman functions completely general. Observe
that in this way the Tolman functions are determined by the spectrum of
perturbations to the original FRW model.

We shall focus on simulating in our spherically symmetric model the results
of concrete cosmological observations such as the redshift - luminosity
distance and the number count - redshift relationships. Since these
observations rely on information carried by light, the analysis centers on
the study of the null geodesics in the model. We device a perturbative
method based on the Hamilton - Jacobi equation for a time like geodesic and
obtain the null geodesic as a limiting case.

We apply our scheme to study two situations. In the first one we consider an
observer located in the center of the Universe and evaluate the number count
and luminosity distance vs redshift relationships. For the first
cosmological test we assume the simplifying hypothesis that the luminosity
of galaxies does not evolve, which is justified in view of the fact that the
redshifts we consider are small ($z<0.08$). For larger redshifts however,
the evolution of luminosity ought to be taken into account \cite{Campos}.

In the second situation we consider an observer located away from the center
and study the anisotropies in the cosmic background radiation due to the
presence of perturbations in the photon path, from the last scattering
surface to the observer's position.

Throughout the paper a unique matter component of the Universe is assumed.

We begin our discussion by briefly reviewing the theory of linearized scalar
adiabatic perturbations to a FRW background, the main features of the Tolman
solution to the Einstein equations and the matching procedure. In section 
{\bf III} we review the cosmological observations $N$ vs $z$ and $d_\ell $
vs $z$ and write their form for a Tolman Universe. In section {\bf IV} we
begin by sketching our method, we perform the explicit calculations for a
centered observer and find the general expression for the mentioned
observational tests. In section {\bf V} we do the same as in section {\bf IV}
but for a non centered observer and evaluate the anisotropies in the CMBR
temperature due to the different paths of the photon from the last
scattering surface to the observer's position. To obtain an idea of the kind
of results to be expected from concrete perturbation profiles, in section 
{\bf VI} we apply our formulae to a particular Tolman Universe,
corresponding to a scale invariant spectrum of perturbations. Finally we
discuss our results in section {\bf VII}.

The overall conclusion is that the method we propose affords a simple way to
model the growth of nonlinear structures such as large scale voids, and may
be used profitably in testing competing theories of primordial fluctuation
generation.

\section{Matching a FRW Universe with Perturbations to a Tolman Solution}

As stated in the Introduction, the accepted framework for structure
formation consists in non linear growth of perturbations during the matter
dominated era of the Universe. The origin of these perturbations are the
zero point fluctuations of the scalar field responsible for driving
Inflation. As the Universe inflates, the wavelengths of the modes that build
fluctuation grow and become bigger than the Hubble length, leaving the
region where microphysics can act. At this stage, the amplitude of each mode
is frozen and so remains until the mode reenters the Hubble sphere. As long
as a mode is outside the horizon its evolution can be studied within linear
theory. We therefore can have an accurate prediction of the amplitudes of
all those modes whose wavelengths are equal to or bigger than the Hubble
length of the considered epoch. Substantial growth, moreover, occurs only
during the matter dominated epoch of the Universe (modes that reenter the
horizon during the radiation dominated epoch grow logarithmically at most 
\cite{meszaros}).

In this context, we will characterize the Tolman Universe by matching it to
a flat Friedmann Robertson Walker Universe with perturbations, at the epoch
of equilibrium between the matter and radiation energy densities.

To simplify the matching procedure we only consider modes bigger than (or at
most equal to) the Hubble size at that epoch. This is certainly justified in
hot dark matter models, where smaller fluctuations are washed out by Landau
damping and free streaming; in more general models a consideration of the
transfer function would be called for. We shall neglect decaying
perturbations to a FRW model, but otherwise do not assume any particular
spectrum for the perturbations. 

We shall also assume that light traces matter, i.e. a Universe with only one
matter component.

\subsection{Scalar Perturbations in a Flat FRW Universe}

The most general metric tensor that describes spherically symmetric scalar
perturbations in a spatially flat FRW Universe is \cite{BranFelMu92} 
\begin{equation}
\delta g_{\mu \nu }=\left( 
\begin{array}{cccc}
2\phi & -aB^{\prime } & 0 & 0 \\ 
-aB^{\prime } & 2a^2\left[ \psi -E_{\mid rr}\right] & 0 & 0 \\ 
0 & 0 & 2a^2\left[ \psi -E_{\mid \theta \theta }\right] & 0 \\ 
0 & 0 & 0 & 2a^2\left[ \psi -E_{\mid \varphi \varphi }\right] \sin ^2\theta
\end{array}
\right)  \label{pert-frw}
\end{equation}
where $\phi $, $\psi $, $B$, $E$ are scalar functions of $r$ and $t$, a
prime means partial derivative with respect to $r$, a dot means partial
derivative with respect to the cosmic time $t$ and a bar means spatial
covariant derivative. $a=(t/t_0)^{2/3}$ is the scale factor for an
unperturbed FRW Universe, where the density is critical, $\Omega _0=1$.

We choose $t_0$ as the age of the Universe at the time of equilibrium
between the matter and radiation energy contents of the Universe ($%
(1+z)_{eq}=2.32\times 10^4$), $t_0\simeq 5.65\times 10^{-4}h^{-1}Mpc$ \cite
{Kolb-Turner} (the speed of light, $c=1$, and $h$ is the fudge factor for
the Hubble constant, $H=100h\,\,kms^{-1}Mpc^{-1}\rightarrow 3\times
10^{-4}h\,\,Mpc^{-1}$ in natural units).

The theory of linearized scalar perturbations is treated extensively in Ref. 
\cite{BranFelMu92}. In the longitudinal gauge ($B=E=0$), in which the
equations take the same form as in the gauge invariant formulation, with a
stress energy tensor whose spatial part is diagonal (which gives $\psi =\phi 
$), if we consider adiabatic perturbations with wavelengths bigger than the
Hubble length, the linearized Einstein equations can be written as a
conservation law for the quantity 
\begin{equation}
\zeta =\frac{2\dot \psi }{3H}+\frac{5\psi }3  \label{zeta-cons}
\end{equation}
where $H=\dot a/a$. For a matter dominated Universe, the solution is 
\begin{equation}
\psi (r,t)=-\frac 35\frac{t_0^{8/3}C_1(r)}{t^{5/3}}+C_2(r)  \label{psi-grow}
\end{equation}
whereby $\zeta =C_2(r)$ follows. The relationship between $\zeta $ and $\psi 
$ can also be written in terms of the respective Fourier modes as 
\begin{equation}
\zeta _k=\frac{2\dot \psi _k}{3H}+\frac{5\psi _k}3  \label{zeta-k}
\end{equation}
$\zeta _k=\delta \rho _k/\rho $ represents the departure from homogeneity in
the density, caused by that mode of the perturbation, and is directly
related to the anisotropies of the microwave background radiation, $\delta
T/T$ (\cite{Borner},\cite{Peebles93}); $k^3\zeta _k$ is related to the
excess mass $\delta M/M$ over a region of volume $k^{-3}$. From COBE
measurements we know that $k^3\zeta _k\simeq 10^{-5}g(k)$, where $g(k)$ is a
smooth, order one function of $k$ \cite{Smoot}.

\subsection{The Tolman Universe}

The Tolman metric for the motion of spherically symmetric dust \cite
{Land-Lifsh} is 
\begin{equation}
ds^2=dt^2-\frac{R^{\prime 2}(r,t)}{f^2(r)}dr^2-R^2(r,t)\left( d\theta
^2+\sin ^2\theta d\varphi ^2\right)  \label{ds2-tolman}
\end{equation}
The Einstein field equations reduce to a single equation \cite{Land-Lifsh} 
\begin{equation}
\dot R^2(r,t)-\frac{F(r)}{R(r,t)}=f^2(r)-1  \label{eq-eins-tol}
\end{equation}
where $R(r,t)>0$ and $f^2(r)$ and $F(r)$ are two arbitrary functions. $F(r)$
can be interpreted as the mass contained in a sphere, and $f^2(r)-1$ as the
mechanical energy of a shell, both of radius $r$.

If $f^2(r)-1$, the Tolman equation can be solved explicitly to yield 
\begin{equation}
R(r,t)=\left[ \frac 94F(r)\right] ^{1/3}(t-t_0)^{2/3}  \label{tolman-flat}
\end{equation}
Otherwise, we do not have an explicit solution, but the relationship of $R$
to $t$ can still be given in terms of a parameter $\eta $ as follows:

$f^2>1$%
\begin{equation}
R(r,\eta )=\frac{F(r)}{2[f^2(r)-1]}(\cosh \eta -1)  \label{R-tol-open}
\end{equation}
\begin{equation}
t(r,\eta )=\frac{F(r)}{2[f^2(r)-1]^{3/2}}(\sinh \eta -\eta )
\label{t-tol-open}
\end{equation}

$f^2<1$%
\begin{equation}
R(r,\eta )=\frac{F(r)}{2[f^2(r)-1]}(1-\cos \eta )  \label{R-tol-closed}
\end{equation}
\begin{equation}
t(r,\eta )=\frac{F(r)}{2[f^2(r)-1]^{3/2}}(\eta -\sin \eta )
\label{t-tol-closed}
\end{equation}

If $f^2-1\sim kr^2$ and $F(r)\sim r^3$, the Tolman solution reduces to a
dust dominated FRW Universe. In this limit, $\eta $ becomes the well known
``conformal'' time. Observe that, in this case $f^2(r)$ equal to, larger, or
less than 1, corresponds to flat, open or closed spatial sections
respectively. The Tolman solution allows for a third arbitrary function of $%
r $, namely, an additive constant in the right hand sides of eqs. (\ref
{t-tol-open}) and (\ref{t-tol-closed}). However, we shall make the
simplifying assumption of disregarding this extra freedom, which can be
shown to be of little relevance to the results below.

\subsection{Matching FRW with Perturbations to Tolman}

Let us now rewrite the perturbed FRW Universe of subsection {\bf A} in the
shape of a Tolman metric. To achieve this, we perform an infinitesimal
change of coordinates $t_{Tolman}=t_{FRW}+{\cal T}(r,t)$, $%
r_{Tolman}=r_{FRW}+{\cal R}(r,t)$, and match $g_{\mu \nu }^{Tolman}=g_{\mu
\nu }^{FRW}$ and $K_{ij}^{Tolman}=K_{ij}^{FRW}$ ($K_{ij}$ the extrinsic
curvature of a spacelike surface) at the surface of matter - radiation
equilibrium. Starting from the longitudinal gauge we get 
\begin{equation}
{\cal T}(r,t)=-\frac 9{10}\frac{t_0^{8/3}C_1(r)}{t^{2/3}}-C_2(r)t
\label{t-Tol-FRW}
\end{equation}

\begin{equation}
{\cal R}(r,t)=\frac 9{10}\frac{t_0^4C_1^{^{\prime }}(r)}t-\frac 32%
t_0^{4/3}C_2^{^{\prime }}(r)t^{2/3}  \label{R-Tol-FRW}
\end{equation}

\begin{equation}
f^2(r)=1-\frac{10}3rC_2^{^{\prime }}(r)  \label{fchic}
\end{equation}

\begin{equation}
F(r)=\frac{4r^3}{9t_0^2}[1-5C_2(r)]  \label{fgran}
\end{equation}

We obtain the same expressions for $f^2(r)$ and $F(r)$ if we start from the
synchronous gauge, ${\cal T}$ and ${\cal R}$ being much simpler. We also
note that the expressions for $f^2(r)$ and $F(r)$ depend only on $C_2(r)$,
the growing mode \cite{Hell-Lake1}. In general the $r$ that appears in the
FRW expressions is a generic length scale while the one in the Tolman
expressions is the coordinate distance from the center of the Universe. In
our matching procedure, the FRW coordinate becomes a radial coordinate. This
is not contradictory in view of the fact that FRW Universes can be
considered a special case of the Tolman solution.

The relevant growth of perturbations starts at $t_0$, the beginning of the
matter dominated epoch of the Universe. Modes that reentered the horizon
before this time at most grow logarithmically, and short wavelength modes
are suppressed through free streaming and Landau damping \cite{Peebles93}.
Therefore at $t_0$ we can write the perturbation as a superposition of all
modes with wave number smaller than a certain cut-off $k_0$, corresponding
to the commoving wavelength of the smallest structure that could be formed.
We then have 
\begin{equation}
C_2(r)=\int\limits_0^{k_0}k^2dk\zeta _k\frac{\sin kr}{kr}  \label{C2-superp}
\end{equation}
where we took the Fourier expansion in spherically symmetric plane waves due
to the symmetry of the Universe. $\zeta _k$ is the spectrum of primeval
fluctuations, and is to be specified on a $t=const.$ surface.

Let us write

\begin{equation}
\zeta _k=\beta (k_0)\xi (u,k_0)  \label{zetak-def1}
\end{equation}
where $u=k/k_0$ and $\xi $ is an order one function. Then $C_2(r)$ will take
the form

\begin{equation}
C_2(r)\sim \gamma (k_0)\xi (x)  \label{C2-integ}
\end{equation}
where $x=k_0r$ and 
\begin{equation}
\xi (x)=\frac 1x\int\limits_0^1du\,u^2\xi (u)\frac{\sin xu}u  \label{Xi(x)}
\end{equation}
The form of $\gamma (k_0)$ depends on the functional form for $\xi (k)$,
which for the moment we leave completely general. In Section {\bf VI} we
will find concrete form for $\gamma (k_0)$ by choosing a particular spectrum
for the perturbations.

>From now on we will employ $x$, rather than $r$ itself, as radial
coordinate. Of course, this coordinate change does not affect the form of
the Tolman metric. The expressions for $F(r)$ and $f^2(r)$ become

\begin{equation}
F(x)=\frac{4x^3}{9t_0^2k_0^3}[1-5\gamma \xi (x)]  \label{fgran(x)}
\end{equation}

\begin{equation}
f^2(x)=1-\frac{10}3\gamma x\xi ^{\prime }(x)  \label{fchic(x)}
\end{equation}

\section{Cosmological observations in a Tolman Universe}

Having at our disposal an exact solution of Einstein equations, we should in
principle be able to follow the nonlinear evolution of structures in the
Universe in all detail. Our observations, on the other hand, are restricted
to that part of the Universe that we can ``see'', that is, to the past
directed light cone with vertex at Earth and now. Therefore in order to make
contact with directly observable magnitudes, we must first discuss how these
quantities are related to the parameters in the Tolman metric, and also how
they are manifest to us as we look back to the decoupling era from our
present location in space time.

Of the possible observations of cosmological relevance, we have chosen to
discuss in detail the dependence of luminosity distances and number counts
with redshift, and the anisotropy in the temperature of the cosmic microwave
background. In this section we shall introduce these concepts, discuss the
shape of the past directed null geodesics in the Tolman Universe, and derive
the main formulae to reconstruct the predictions of these models concerning
redshift surveys and the Hubble law.

\subsection{Luminosity distances}

Let us begin our discussion introducing a notion of distance to a
cosmological object which can be inferred from Earth - based observations.
Let $d\Omega _o$ be the solid angle subtended by a bundle of null geodesics
diverging from the observer, and let $dS_o$ be the cross - sectional area of
this bundle at some point. Then the observer area distance $D_o$ of this
point from the observer is defined by \cite{Ellis} 
\begin{equation}
dS_o=D_o^2d\Omega _o  \label{oad}
\end{equation}

Thus we can find $D_o$ if we can measure the solid angle subtended by some
object whose cross - sectional area can be found from astrophysical
considerations. $D_o$ is the same as the corrected luminosity distance \cite
{Ellis} and also the same as the angular diameter distance \cite{Weinberg72}%
. For a Friedmann - Robertson - Walker Universe the expression for $D_o$ is 
\begin{equation}
D_o=a(t_s)\,x  \label{adfrw}
\end{equation}
where and $t_s$ indicates time of source emission. For the Tolman Universe
we have 
\begin{equation}
D_o=R(x,t_s)  \label{toad}
\end{equation}

The luminosity distance{\sl \ }$d_\ell $ is defined as 
\begin{equation}
d_\ell =\sqrt{\frac{F_s}F}  \label{dl}
\end{equation}
where $F_s$ is the flux of the source measured in its neighborhood and $F$
is the observed flux. The corrected luminosity distance $d$ is defined as 
\begin{equation}
d=\frac{d_\ell }{(1+z)^2}  \label{cld}
\end{equation}
where $z$ is the redshift of the source measured by the observer. By the
reciprocity theorem \cite{Ellis}, we have $d=D_o$, and the luminosity
distance $d_\ell $ for the Tolman Universe becomes 
\begin{equation}
d_\ell =R(x,t)(1+z)^2  \label{tld}
\end{equation}

\subsection{Number counts}

One of the cosmological quantities used to study the large scale structures
in the Universe is the number count of a certain structure (galaxy or
cluster) as a function of the redshift $z$. The starting point to calculate
this quantity is the number of sources in a section of a bundle of past null
geodesics \cite{Ellis}, which for a Tolman Universe reads\cite{Ribeiro} 
\begin{equation}
dN=4\pi \nu \frac{R^{/}R^2}fdx  \label{dnct}
\end{equation}
$\nu $ being the number density of structures\cite{Land-Lifsh} 
\begin{equation}
\nu =\frac{F^{/}(x)}{16\pi M_sR^{/}R^2}  \label{n-dens-Tol}
\end{equation}
and where $M_s$ is the rest mass of the structure. The number of sources
which lie at radial coordinate distances less than $x$ as seen by an
observer at $x=0$ is \cite{Ribeiro} 
\begin{equation}
N(x)=\frac 1{4M_s}\int\limits_{LC}dx\frac{F^{/}(x)}{f(x)}  \label{nct}
\end{equation}
where the integration is made along the light-cone, $LC$, parametrized by $x$%
. For an almost spatially flat Universe we can consider that this number
(omitting the mass-dependent prefactor) is 
\begin{equation}
N(x)=F(x)  \label{ncft}
\end{equation}
This quantity ought to be evaluated along a past null geodesic.

\subsection{Redshift}

The redshift of a source as measured by an observer is defined in terms of
frequencies by 
\begin{equation}
1+z=\frac{\nu _{emitter}}{\nu _{observer}}  \label{defz}
\end{equation}
where $\nu _{emitter}$ is the intrinsic frequency of the source and $\nu
_{observer}$ is the value of that frequency detected by the observer. If $%
\lambda $ is an (affine) parameter along the null geodesic between source
and observer and $t$ is the cosmological time, (\ref{defz}) can be written
as 
\begin{equation}
1+z={\frac{(dt/d\lambda )|_x}{(dt/d\lambda )|_{x=0}}}  \label{parz}
\end{equation}
where $x$ denotes the position of the source and where we take the observer
located at $x=0$. We shall discuss further this formula in next subsection.

\subsection{Anisotropies in the CMBR temperature}

In the Tolman Universe, when we move the observer away from the center of
symmetry, the observations that she/he performs will be anisotropic. The
physical system on which the most sensitive analysis of anisotropies are
being performed is the CMBR. In the case of the Tolman Universe there are
two sources of anisotropies: one related to the fluctuations in the density
at the surface of last scattering, $\bigtriangleup T/T\simeq (1/3)[\Delta
\delta /\bar \delta ]_{dec}$, and the other related to the different paths
of the photons from the last scattering surface to the non-centered
observer, $\bigtriangleup T/T\simeq [z(\theta )-\bar z]_{dec}$. We will
calculate both anisotropies for our model and compare one to the other.

\section{Radial Geodesics in the Tolman Universe}

Having derived the generic expressions for luminosity distances and number
counts in terms of Tolman coordinates and the $F$ and $f$ functions, the
only remaining step is to find the evolution of these coordinates as we
travel towards the past along a null geodesic. To this end, it is convenient
to think of a null geodesic as the limiting case of a time like geodesic.
Time like geodesics have a natural parametrization in terms of proper time
or some other affine parameter, and the form of the geodesic can be found by
integrating a Hamiltonian system. We obtain a null geodesic by further
imposing the constraint that the Hamiltonian be null; the affine
parametrization of the time like geodesics then induces a well defined
parametrization on the null geodesic.

We can write the action for radial geodesics \cite{Gravitation} in Tolman
Universe as

\begin{equation}
S=\frac 12\int d\lambda \left\{ -\left( {\frac{dt}{d\lambda }}\right) ^2+{\ 
\frac{R^{\prime }{}^2(x,t)}{f^2(x)}}\left( {\frac{dx}{d\lambda }}\right)
^2+R^2(x,t)\left[ \left( \frac{d\theta }{d\lambda }\right) ^2+\sin ^2\theta
\left( \frac{d\varphi }{d\lambda }\right) ^2\right] \right\}  \label{geod}
\end{equation}
where $\lambda $ is an affine parameter for timelike and spacelike
geodesics. The canonically conjugated momenta and the Hamiltonian are

\begin{equation}
P_t=-\frac{dt}{d\lambda }  \label{pt}
\end{equation}
\begin{equation}
P_r=\frac{R^{/2}(x,t)}{f^2(x)}\frac{dx}{d\lambda }  \label{pr}
\end{equation}
\begin{equation}
P_\theta =R^2(x,t)\frac{d\theta }{d\lambda }  \label{ptita}
\end{equation}
\begin{equation}
P_\varphi =R^2(x,t)\sin ^2\theta \frac{d\varphi }{d\lambda }  \label{pfi}
\end{equation}
\begin{equation}
H=-P_t^2+\frac{f^2(x)}{R^{/2}(x,t)}P_r^2+\frac 1{R^2(x,t)}P_\theta ^2+\frac 1%
{R^2(x,t)\sin ^2\theta }P_\varphi ^2  \label{ham}
\end{equation}
For $H\rightarrow 0$ we have a null geodesic , and in this case the redshift
is given by (cfr. eq. (\ref{parz}) )

\begin{equation}
1+z={\frac{(dt/d\lambda )|_x}{(dt/d\lambda )|_{x=0}}}={\frac{P_t|_x}{%
P_t|_{x=0}}}  \label{z-tolman}
\end{equation}
We choose the parameter $\lambda $ so that ${P_t|_{x=0}=1}$, then ${P_t|_x}$
is directly the redshift at that location.

In order to proceed with the integration of these equations, we should know
the expression for $R$ as a function of $t$ and $x$, which entails solving
the parametric equations (\ref{R-tol-open}) and (\ref{t-tol-open}) or (\ref
{R-tol-closed}) and \ref{t-tol-closed}), with known $F$ and $f$. Obviously
this cannot be done in closed form unless in the trivial case $f\equiv 1$.
Nevertheless we can solve those equations perturbatively, developing the
hyperbolic (or trigonometric) functions around the origin ($t\rightarrow 0$, 
$x\simeq 0$), the corresponding expressions for $R$, $R^{-2}$ and $R^{\prime
-2}$ can be read in Appendix {\bf A}.

Let us consider the evolution of the perturbations in a neighborhood of the
origin. We have to develop the functions of $x$ around $x\simeq 0$. For the
function $F$ we approximate it by 
\begin{equation}
F(x)\sim \frac{4x^3}{9t_0^2k_0^3}  \label{fgran-approx}
\end{equation}
since $\gamma \sim O(10^{-5})$. From eqs. (\ref{t-tol-open}) and (\ref{fchic}%
) we obtain 
\begin{equation}
\sinh \eta -\eta \simeq 2t\frac{\left[ 10\gamma x\xi ^{\prime }(x)/3\right]
^{3/2}}{4x^3/9t_0^2k_0^3}  \label{tdevel}
\end{equation}
(of course the same argument applies in the trigonometric case) To achieve a
non null value of $\eta $ at the origin we have to require that $x\xi
^{\prime }(x)$ goes to $0$ like $ax^2$, where $a$ is some constant of order
one. In this case eq. (\ref{tdevel}) yields 
\begin{equation}
\eta _{today}=3(k_0^3t_0^2t_{today})^{1/3}\sqrt{\frac{10\gamma a}3}
\label{etadef}
\end{equation}

Since $\eta $ decreases as we move down the geodesic, we find that
throughout we remain in a neighborhood of $\eta =0$. We are led then to an
approximation scheme whereby, in the $m-th$ order approximation, we replace
the hyperbolic (or trigonometric) functions in the parametric equations by
the first nontrivial $m$ terms in their Taylor expansion around the origin.
For example, at leading order we approximate $\sinh \eta -\eta \sim \eta
^3/6 $; at next to leading order we retain also the $\eta ^5$ term, and so
on. In other words, we are neglecting terms of order $\eta ^{2m+2}/(2m+2)!$
against those of order $\eta ^{2m}/(2m)!$. This approximation is seen to be
valid if $\eta ^2\ll (2m+1)(2m+2)$. $\eta _{today}$ is a measure of the non
linear evolution of the perturbation: higher $\eta _{today}$ means higher
non linearity and therefore more orders are to be considered in the
development of the hyperbolic (trigonometric) expressions of $t$ and $R$.
>From eq. (\ref{etadef}) we see that increasing $\eta _{today}$ means
increasing $k_0$, i.e. considering a perturbation characterized by a smaller
wavelength and that consequently is in a more non linear stage of evolution
today.

\subsection{Leading Order Approximation}

For radial geodesics the Hamiltonian reads 
\begin{equation}
H=-P_t^2+\frac{f^2(x)}{R^{/2}(x,t)}P_x^2  \label{Hrad}
\end{equation}
Starting from eqs. (\ref{R-tol-open}) and (\ref{t-tol-open}) and taking the
lowest order in the development of the hyperbolic functions ($\eta \ll 1$)
we have 
\begin{equation}
\eta ^3\simeq \frac{12\left[ f^2(x)-1\right] ^{3/2}t}{F(x)}  \label{eta-cero}
\end{equation}
and therefore 
\begin{equation}
R(x,t)\simeq \frac{12^{2/3}F^{1/3}(x)\,t^{2/3}}4  \label{R-cero}
\end{equation}
The Hamilton-Jacobi equation for the characteristic function is 
\begin{equation}
-\left( \frac{\partial W}{\partial t}\right) ^2+\frac 1{v_0^2(x)t^{4/3}}%
\left( \frac{\partial W}{\partial x}\right) ^2=-E  \label{H-J-cero}
\end{equation}
with 
\[
v_0(x)=\frac{F^{\prime }(x)}{12^{1/3}F^{2/3}(x)f(x)} 
\]
and where we used as the principal function $S=W+\lambda E$. We separate
variables by writing $W=T(t,p,E)+pL(x)$, with 
\begin{equation}
\frac{dL}{dx}=v_0(x)  \label{dL/dx-rad}
\end{equation}
\begin{equation}
\frac{dT}{dt}=\sqrt{E+\frac{p^2}{t^{4/3}}}  \label{dT/dt-rad}
\end{equation}
As we are interested only in a neighborhood of $E=0$, we can expand eq. (\ref
{dT/dt-rad}) 
\begin{equation}
\frac{dT}{dt}\simeq \frac p{t^{2/3}}+\frac 12E\frac{t^{2/3}}p-\frac 18E^2%
\frac{t^2}{p^3}+\cdots  \label{dT/dt-rad-dev}
\end{equation}
which can be integrated to give 
\begin{equation}
T\simeq 3pt^{1/3}+\frac 3{10}E\frac{t^{5/3}}p-\frac 1{24}E^2\frac{t^3}{p^3}%
+\cdots  \label{T-rad-dev}
\end{equation}

If we define new canonical variables $\tau $, $\rho $ conjugated to $E$ and $%
p$ respectively, we have from the characteristic function $W$ that 
\begin{equation}
\tau =\frac{\partial W}{\partial E}=\frac 3{10}\frac{t^{5/3}}p-\frac 1{12}E%
\frac{t^3}{p^3}+\cdots  \label{tau-rad-def}
\end{equation}
\begin{equation}
\rho =\frac{\partial W}{\partial p}=L(x)+3t^{1/3}-\frac 3{10}E\frac{t^{5/3}}{%
p^2}+\frac 18E^2\frac{t^3}{p^4}-\cdots  \label{rho-rad-def}
\end{equation}
from where 
\begin{equation}
L(x)=\rho -3t^{1/3}+\frac 3{10}E\frac{t^{5/3}}{p^2}-\frac 18E^2\frac{t^3}{p^4%
}+\cdots  \label{L(x)}
\end{equation}
We can write the old momentum $P_t$ in terms of the new momenta as 
\begin{equation}
P_t=\frac{\partial W}{\partial t}=\sqrt{E+\frac{p^2}{t^{4/3}}}=\frac p{%
t^{2/3}}+\frac 12E\frac{t^{2/3}}p-\frac 18E^2\frac{t^2}{p^3}+\cdots
\label{Pt-rad-def}
\end{equation}
The new Hamiltonian is $H=-E$. The equations of motion are 
\begin{eqnarray}
\dot \tau &=&\partial H/\partial E=-1  \label{eqmot-rad} \\
\dot E &=&\dot p=\dot \rho =0  \nonumber
\end{eqnarray}
where a dot means $d/d\lambda $. As $p=const.$ we can choose it as 
\[
p=t_{today}^{2/3} 
\]
so that 
\begin{equation}
P_t|_{E=0}=(1+z)=\left( \frac{t_{today}}t\right) ^{2/3}  \label{z-FRW}
\end{equation}
therefore when $t=t_{today}$ we have $z=0$. We read the spatial coordinate
from eq. (\ref{rho-rad-def}) and as also $\rho =const.$ we take it as 
\[
\rho =3t_{today}^{1/3} 
\]
therefore from eq. (\ref{rho-rad-def}) we have 
\[
L(x)|_{E=0}=3t_{today}^{1/3}-3t^{1/3} 
\]
We can use this last equation to find an expression for $F(x)$. From Eq.(\ref
{dL/dx-rad}) and the definition of $v_0(x)$ we have 
\begin{equation}
L(x)=\frac{12^{2/3}}4\left\{ \frac{F^{1/3}(x)}{f(x)}+\int dx\frac{%
F^{1/3}(x)f^{\prime }(x)}{f^2(x)}\right\}  \label{L(x)(F(x))}
\end{equation}
As the second term is $O(\gamma )$ we can neglect it. So for $f(x)\sim 1$ we
have 
\begin{equation}
F^{1/3}(x)=\frac 4{12^{2/3}}L(x)=(12t_{today})^{1/3}\left[ 1-\frac{t^{1/3}}{%
t_{today}^{1/3}}\right]  \label{F(L(x))}
\end{equation}
Finally, recalling Eq. (\ref{fgran-approx}) the equation for the geodesic
becomes 
\begin{equation}
x=x_0\left[ 1-\frac{t^{1/3}}{t_{today}^{1/3}}\right]  \label{rad-geod}
\end{equation}
where $x_0=3(k_0^3t_0^2t_{today})^{1/3}=(10\gamma a/3)^{-1/2}\eta _{today}$.

According to eqs. (\ref{F(L(x))}), (\ref{R-cero}), (\ref{z-FRW}) and (\ref
{tld}) the luminosity distance reads 
\begin{equation}
d_\ell =3t_{today}\left[ (1+z)-\sqrt{1+z}\right]  \label{dl-O0-C}
\end{equation}
where we see that for small $z$, $d_\ell \simeq H_0^{-1}z$, $%
H_0=2/3t_{today} $ as expected for a flat, matter dominated FRW Universe.

\subsection{Next to Leading Order Approximation}

Now the Hamilton-Jacobi equation reads 
\begin{equation}
-\left( \frac{\partial W}{\partial t}\right) ^2+\left( \frac{\partial W}{%
\partial x}\right) ^2\frac 1{v_0^2(x)t^{4/3}}\left[ 1-\frac 1{10}%
v_1(x)t^{2/3}\right] =-E  \label{HamJac-O1}
\end{equation}
where 
\begin{equation}
v_1(x)=\frac{12^{2/3}}{[F^{1/3}(x)]^{\prime }}\left( \frac{f^2(x)-1}{%
F^{1/3}(x)}\right) ^{\prime }=-\frac{\eta _{today}^2}{at_{today}^{2/3}}\xi
^{\prime \prime }(x)  \label{v1(x)}
\end{equation}
$E$ and $p$ are no longer constants of motion because in terms of them 
\begin{equation}
H=-E-p^2\frac{v_1(x)}{10t^{2/3}}  \label{Ham-O1}
\end{equation}
We therefore perform a further canonical transformation to variables $E_1$, $%
p_1$, $\rho _1$, $\tau _1$ so that $H=-E_1$. The generating function is now 
\begin{equation}
W_1=E_1\tau +p_1\rho -p_1^2G_1(E_1,p_1,\rho ,\tau )  \label{W1}
\end{equation}
The old momentum $E$ becomes 
\begin{equation}
E=\frac{\partial W_1}{\partial \tau }=E_1-p_1^2\frac{\partial G_1}{\partial
\tau }  \label{E-O1}
\end{equation}
so to obtain $H=-E_1$ we demand 
\begin{equation}
\frac{\partial G_1}{\partial \tau }=\frac{v_1(x)}{10t^{2/3}}  \label{G1(def)}
\end{equation}
We can write formally 
\begin{equation}
G_1=\frac 1{10}\int\limits_{\tau _{today}}^\tau d\tau ^{\prime }\frac{%
v_1[x(E_1,p_1,\rho ,\tau ^{\prime })]}{t^{2/3}(E_1,p_1,\tau ^{\prime })}
\label{G1(form.)}
\end{equation}
Under the integration sign we may use the zeroth order relationships among
the different variables, namely $E_1\sim E$, $p_1\sim p$, and 
\begin{equation}
\tau =\frac{dT}{dE}=\frac d{dE}\left( \int\limits_{t_{today}}^tdt^{\prime }%
\sqrt{E+\frac{p^2}{t^{4/3}}}\right) \rightarrow d\tau =\frac 1{2p}\frac{%
t^{2/3}}{\sqrt{1+\frac{Et^{4/3}}{p^2}}}dt  \label{dtau(dt)}
\end{equation}
so 
\begin{equation}
G_1=\frac 1{20p_1}\int\limits_{t_{today}}^t\frac{dt^{\prime }}{\sqrt{1+\frac{%
E_1t^{4/3}}{p_1^2}}}v_1[x(E_1,p_1,\rho ,t^{\prime })]  \label{G1(form.)(t)}
\end{equation}
where $t=t(\tau ,E_1,p_1)$. Let us now compute the corrections to the
several quantities introduced so far.

\subsubsection{Corrections to $E$}

The value of $E$ valid to this order follows immediately from eqs. (\ref
{E-O1}) and (\ref{G1(def)}) plus the Hamiltonian constraint $E_1=0$, leading
to 
\[
E=-\frac{p_1^2}{10}\frac{v_1(x)}{t^{2/3}} 
\]
where $t=t(\tau _1,E_1=0,p_1)$, $x=x(\tau _1,\rho _1,E_1=0,p_1)$.

\subsubsection{Corrections to $p$}

The value of the old momentum $p$ follows from 
\[
p=\frac{\partial W_1}{\partial \rho }=p_1-p_1^2\frac{\partial G_1}{\partial
\rho } 
\]
Eq. (\ref{G1(form.)(t)}) leads to 
\begin{equation}
\frac{\partial G_1}{\partial \rho }=\frac 1{20p_1}\int\limits_{t_{today}}^t%
\frac{dt^{\prime }}{\sqrt{1+\frac{E_1t^{4/3}}{p_1^2}}}v_1^{\prime }(x)\frac{%
\partial x}{\partial \rho }|_{E_1,p_1,t^{\prime }}  \label{dG1/drho}
\end{equation}
where $v_1^{\prime }(x)=dv_1(x)/dx$ and where the integration is to be
performed along a past null geodesic. As $v_1(x)$ is of first order, $%
\partial x/\partial \rho $ may be evaluated from the zeroth order, giving $%
\partial x/\partial \rho =x_0/3t_{today}^{1/3}$. Therefore, setting $E_1=0$
we get 
\begin{equation}
\frac{\partial G_1}{\partial \rho }|_{E_1=0}=\frac 1{20p_1}\frac{x_0}{%
3t_{today}^{1/3}}\int\limits_{t_{today}}^tdt^{\prime }v_1^{\prime }(x)
\label{dG1/drho(E1=0)}
\end{equation}
and 
\begin{equation}
p=p_1\left\{ 1-\frac 1{20}\frac{x_0}{3t_{today}^{1/3}}\int%
\limits_{t_{today}}^tdt^{\prime }v_1^{\prime }(x)\right\}  \label{p-orden1}
\end{equation}

\subsubsection{Corrections to $\rho $ and $\tau $}

The new coordinates are 
\begin{equation}
\rho _1=\frac{\partial W_1}{\partial p_1}=\rho -2p_1G_1-p_1^2\frac{\partial
G_1}{\partial p_1}  \label{rho-orden1}
\end{equation}
\begin{equation}
\tau _1=\frac{\partial W_1}{\partial E_1}=\tau -p_1^2\frac{\partial G_1}{%
\partial E_1}  \label{tau-orden1}
\end{equation}
For $\rho $ we have from (\ref{G1(form.)(t)}) that at $E_1=0$%
\begin{equation}
\frac{\partial G_1}{\partial p_1}=-\frac{G_1}{p_1}+\frac 1{20p_1}v_1(x)\frac{%
\partial t}{\partial p_1}-\frac 1{20p_1}\int\limits_{t_{today}}^tdt^{\prime
}v_1^{\prime }(x)\frac{\partial x}{\partial p_1}|_{E,\rho ,t}
\label{dG1/dp1}
\end{equation}
>From the zeroth order we have $\rho =3t^{1/3}+L(x)$, and consequently $%
\partial x/\partial p=0$ and from $t=(10p\tau /3)^{3/5}$ we get $\partial
t/\partial p|_\tau =3t/5p_1$. Therefore 
\begin{equation}
\rho =\rho _1+\frac{3t}{100}v_1(x)+\frac 1{20}\int\limits_{t_{today}}^tdt^{%
\prime }v_1(x)  \label{rho(O1)}
\end{equation}

Finally for $\tau $ we have (cf. Eq. (\ref{tau-orden1})) 
\begin{eqnarray}
\tau &=&\tau _1+\frac 1{12p_1}v_1(x)\,t^{7/3}+\frac{x_0}{20t_{today}^{1/3}p_1%
}\int\limits_{t_{today}}^tdt\,t^{5/3}\,v_1^{^{\prime }}(x)-\frac 1{4p_1}%
\int\limits_{t_{today}}^tdt\,t^{5/3}\,v_1(x)  \label{tau(O1)}
\end{eqnarray}
where we have used the zeroth order results $\partial t/\partial
E_1|_{E=0}=t^{7/3}/6p^2$ and $\partial x/\partial
E_1|_{E=0}=x_0t^{5/3}/10t_{today}^{1/3}p^2$.

\subsubsection{Redshift, Number Count, Density Contrast and Luminosity
Distance}

Having computed the old canonical coordinates in terms of the new ones, it
is only a matter of substituting them into eqs. (\ref{L(x)}) and (\ref
{Pt-rad-def}) to obtain the formulae for the number counts (through the
auxiliary function $L(x)$, recall Eq. (\ref{F(L(x))})) and redshifts. For $%
L(x)$ we get 
\begin{equation}
L(x)=\rho -3t^{1/3}+\frac 3{10}E\frac{t^{5/3}}{p^2}=\rho _1+\frac 1{20}%
\int\limits_0^tdt^{\prime }v_1(x)-3t^{1/3}  \label{L(x)-ord1}
\end{equation}
Imposing the boundary condition $L(0)=0$, we get 
\begin{equation}
L(x)=3(t_{today}^{1/3}-t^{1/3})+\frac 1{20}\int\limits_t^{t_{today}}dt^{%
\prime }v_1(x)  \label{L(x(t))-O1}
\end{equation}

Substituting the expressions for $E$ and $p$ in Eq. (\ref{Pt-rad-def}) for $%
P_t$, and imposing the boundary condition $P_t=1$ at $t=t_{today}$ we get 
\begin{equation}
P_t=\left( \frac{t_{today}}t\right) ^{2/3}\left\{ 1-\frac 1{20}\left(
t^{2/3}v_1(x)-t_{today}^{2/3}v_1(0)\right) -\frac 1{20p_1}\frac{x_0}{%
3t_{today}^{1/3}}\int\limits_{t_{today}}^tdt^{\prime }v_1^{\prime
}(x)\right\}  \label{Pt(t)-O1}
\end{equation}

Now we must find the explicit expressions for $dt=dt(x)$. From the zeroth
order we have 
\begin{equation}
dt=-\frac{3t_{today}}{x_0}\left( 1-\frac x{x_0}\right) ^2dx\simeq -\frac{%
3t_{today}}{x_0}dx  \label{dt(x)}
\end{equation}
where the last expression is due to the fact that we are interested in a
neighborhood of $x=0$. Replacing this expression and Eq. (\ref{v1(x)}) in
eqs. (\ref{L(x)-ord1}) and (\ref{Pt(t)-O1}) we get 
\begin{equation}
P_t=\left( \frac{t_{today}}t\right) ^{2/3}\left\{ 1-\frac{\eta _{today}^2}{%
20a}\xi ^{\prime \prime }(x)\left[ 1-\left( \frac t{t_{today}}\right)
^{2/3}\right] \right\}  \label{Pt-O1-final}
\end{equation}
\begin{equation}
F^{1/3}(x)=\left( 12t_{today}\right) ^{1/3}\left\{ 1-\left( \frac t{t_{today}%
}\right) ^{1/3}+\frac{\eta _{today}}{20}\sqrt{\frac{10\gamma }{3a}}\xi
^{\prime }(x)\right\}  \label{F(x)-O1-final}
\end{equation}
and in this case the number count is given by $N(x)=F(x)$ (cfr. Eq. (\ref
{nct})). The density contrast is given by eq. (\ref{n-dens-Tol}) omitting
the mass factor and in this case we obtain 
\begin{equation}
\frac{\Delta \delta }{\bar \delta }=\frac{\eta _{today}^2}{20a}\left[ \xi
^{^{\prime \prime }}(x)+2\frac{\xi ^{^{\prime }}(x)}x\right] \left( \frac t{%
t_{today}}\right) ^{2/3}  \label{dencon-c-O1}
\end{equation}
where $\bar \delta =(12\pi t^2)^{-1}$ and where we used $\delta $ for the
energy density in order to avoid confusion with notation. Finally, the
luminosity distance reads 
\begin{equation}
d_\ell =3\frac{t_{today}^{5/3}}{t^{2/3}}\left\{ 1-\frac{t^{1/3}}{%
t_{today}^{1/3}}+\frac{\eta _{today}}{20}\sqrt{\frac{10\gamma }{3a}}\left[
\xi ^{^{\prime }}(x)-2x\xi ^{^{\prime \prime }}(x)\right] \left[ 1-\left( 
\frac t{t_{today}}\right) ^{2/3}\right] \right\}  \label{dl-O1-c}
\end{equation}
where these last three quantities are to be evaluated using $%
x=x_0(1-t^{1/3}/t_{today}^{1/3})$. These expressions are completely general,
the only assumption on the function $\xi (x)$ being that $x\xi ^{^{\prime
}}(x)$ goes to zero as $x^2$. Therefore, given a power spectrum $\zeta _k$
we can find the perturbed background, described by $F(x)$ and $f^2(x)$ at a
given initial time $t_0$ and then evolve it non linearly using the Tolman
equations. However, to properly speak about non linear evolution, we have to
consider higher orders. The procedure to evaluate them is the same as the
used for the next to leading order. In the next subsection we sketch the
evaluation of the second order and give the final results.

\subsection{Second Order Approximation}

In this case the Hamilton - Jacobi equation reads 
\begin{equation}
-\left( \frac{\partial W}{\partial t}\right) ^2+\left( \frac{\partial W}{%
\partial x}\right) ^2\frac 1{v_0^2(x)t^{4/3}}\left[ 1-\frac 1{10}%
v_1(x)t^{2/3}+v_2(x)t^{4/3}\right] =-E  \label{H-J-O2}
\end{equation}
where 
\begin{eqnarray*}
v_2(x) &=&\frac 3{400}\frac{12^{4/3}}{[F^{1/3}(x)]^{^{\prime }2}}\left[ 
\frac{f^2(x)-1}{F^{1/3}(x)}\right] ^{^{\prime }2}+\frac 3{1400}\frac{12^{4/3}%
}{[F^{1/3}(x)]^{^{\prime }}}\left[ \frac{(f^2(x)-1)^2}F\right] ^{^{\prime }}
\\
\ &=&\frac 3{400a^2}\frac{\eta ^4}{t_{today}^{4/3}}\xi ^{^{\prime \prime
}2}(x)+\frac 3{1400a^2}\frac{\eta ^4}{t_{today}^{4/3}}\left( \frac{\xi
^{^{\prime }2}}x\right) ^{^{\prime }}
\end{eqnarray*}
If we replace in (\ref{H-J-O2}) the leading and next to leading order we
find again that $E_1$ and $p_1$ are no longer constants of motion because in
terms of them 
\[
H=-E_1+p_1^2v_2(x) 
\]
We perform a further canonical transformation whose generating function is
again of the form 
\[
W_2=E_2\tau _1+p_2\rho _1-p_2^2G_2(\rho _1,\tau _1,p_2,E_2) 
\]
so the first order momentum $E_1$ becomes 
\[
E_1=\frac{\partial W_2}{\partial \tau _1}=E_2-p_2^2\frac{\partial G_2}{%
\partial \tau _1} 
\]
To obtain $H=-E_2$ we proceed as in the previous order and demand 
\[
\frac{\partial G_2}{\partial \tau _1}=-v_2(x) 
\]
The procedure to calculate the corrections to $p_1$, $\rho _1$ is the same,
although longer, as in the next to leading order case. We therefore show
only the final expressions for $E_1$, $p_1$, $\rho _1$, $P_t$, $F^{1/3}(x)$, 
$\Delta \delta /\delta $ and $d_\ell $. 
\begin{equation}
E_1=p_2^2v_2(x)  \label{E1-C-O2}
\end{equation}
\begin{equation}
p_1=p_2\left\{ 1+\frac{x_0}{6t_{today}^{1/3}}\int\limits_{t_{today}}^tdt%
\,t^{2/3}v_2^{^{\prime }}(x)\right\}  \label{p1-C-O2}
\end{equation}
\begin{equation}
\rho _1=\rho _2-\frac 7{10}\int\limits_{t_{today}}^tdt\,t^{2/3}v_2(x)-\frac 3%
{10}t^{5/3}v_2(x)  \label{rho1-C-O2}
\end{equation}
\begin{eqnarray}
P_t &=&\left( \frac{t_{today}}t\right) ^{2/3}\left\{ 1-\frac{\eta ^2}{20a}%
\xi ^{^{\prime \prime }}(x)\left[ 1-\frac{t^{2/3}}{t_{today}^{2/3}}\right] -%
\frac{3\eta ^4}{2800a^2}\left( \frac{\xi ^{^{\prime }2}(x)}x\right)
^{^{\prime }}\left[ 1-\frac{t^{4/3}}{t_{today}^{4/3}}\right] -\right.
\label{resh-c-O2} \\
&&\left. \ -\frac{3\eta ^4}{800a^2}\xi ^{^{\prime \prime }2}(x)\left[ 1-%
\frac{t^{4/3}}{t_{today}^{4/3}}\right] -\frac{\eta ^4}{800a^2}\xi ^{^{\prime
\prime }2}(x)\left[ \frac{t^{2/3}}{t_{today}^{2/3}}-2\frac{t^{4/3}}{%
t_{today}^{4/3}}\right] -\frac{\eta ^4}{800a^2}\xi ^{^{\prime \prime
}2}(0)\right\}  \nonumber
\end{eqnarray}
\begin{eqnarray}
F^{1/3}(x) &=&(12t_{today})^{1/3}\left\{ 1-\frac{t^{1/3}}{t_{today}^{1/3}}+%
\frac{\eta _{today}}{20}\sqrt{\frac{10\gamma }{3a}}\xi ^{^{\prime
}}(x)+\right.  \nonumber \\
&&+\frac{21\eta _{today}^3}{4000}\sqrt{\frac{10\gamma }{3a^3}}%
\int\limits_0^xdx\xi ^{^{\prime \prime }2}(x)+\frac{3\eta _{today}^3}{2000}%
\sqrt{\frac{10\gamma }{3a^3}}\frac{\xi ^{^{\prime }2}(x)}x-  \label{Fgr-c-O2}
\\
&&-\frac{\eta _{today}^4}{2400a^2}\xi ^{^{\prime \prime }2}(x)\frac{t^{5/3}}{%
t_{today}^{5/3}}+\frac{\eta _{today}^4}{2400a^2}\xi ^{^{\prime \prime }2}(0)+
\nonumber \\
&&\left. +\frac{\eta _{today}^4}{1000a^2}\xi ^{^{\prime \prime }2}(x)\frac t{%
t_{today}}-\frac{\eta _{today}^4}{1000a^2}\xi ^{^{\prime \prime
}2}(0)\right\}  \nonumber
\end{eqnarray}
\begin{eqnarray}
\frac{\Delta \delta }{\bar \delta } &=&\frac{\eta _{today}^2}{20a}\left[ \xi
^{^{\prime \prime }}(x)+2\frac{\xi ^{^{\prime }}(x)}x\right] \left( \frac t{%
t_{today}}\right) ^{2/3}-\frac{\eta _{today}^4}{700a^2}\frac{\xi ^{^{\prime
}2}(x)}{x^2}\left( \frac t{t_{today}}\right) ^{4/3}+  \nonumber \\
&&+\frac{\eta _{today}^4}{140a^2}\frac{\xi ^{^{\prime }}(x)\xi ^{^{\prime
\prime }}(x)}x\left( \frac t{t_{today}}\right) ^{4/3}+\frac{\eta _{today}^4}{%
400a^2}\frac{\xi ^{^{\prime }2}(x)}{x^2}\left[ 4\left( \frac t{t_{today}}%
\right) ^{4/3}-3\left( \frac t{t_{today}}\right) ^{2/3}\right] +
\label{decon-C-O2} \\
&&+\frac{\eta _{today}^4}{400a^2}\xi ^{^{\prime \prime }2}(x)\left[ \left( 
\frac t{t_{today}}\right) ^{4/3}-\left( \frac t{t_{today}}\right)
^{2/3}\right] -\frac{\eta _{today}^4}{200a^2}\frac{\xi ^{^{\prime }}(x)\xi
^{^{\prime \prime }}(x)}x\left( \frac t{t_{today}}\right) ^{2/3}  \nonumber
\end{eqnarray}
\begin{eqnarray}
d_\ell &=&3\frac{t_{today}^{5/3}}{t^{2/3}}\left\{ \left( 1-\frac{t^{1/3}}{%
t_{today}^{1/3}}\right) +\frac{\eta _{today}}{20}\sqrt{\frac{10\gamma }{3a}}%
\left[ \xi ^{^{\prime }}(x)-2x\xi ^{^{\prime \prime }}(x)\right] \left( 1-%
\frac{t^{2/3}}{t_{today}^{2/3}}\right) +\right. \\
&&+\frac{21\eta _{today}^3}{4000}\sqrt{\frac{10\gamma }{3a^3}}%
\int\limits_0^xdx\xi ^{^{\prime \prime }2}(x)-\frac{\eta _{today}^3}{400}%
\sqrt{\frac{10\gamma }{3a^3}}x\xi ^{^{\prime \prime }2}(x)\left( 2-3\frac{%
t^{4/3}}{t_{today}^{4/3}}\right) -  \nonumber \\
&&-\ \frac{\eta _{today}^3}{1400}\sqrt{\frac{10\gamma }{3a^3}}\xi ^{^{\prime
}}(x)\xi ^{^{\prime \prime }}(x)\left( 13-14\frac{t^{2/3}}{t_{today}^{2/3}}+%
\frac{t^{4/3}}{t_{today}^{4/3}}\right) +\   \label{dl-c-O2} \\
&&+\frac{\eta _{today}^3}{2800}\sqrt{\frac{10\gamma }{3a^3}}\frac{\xi
^{^{\prime }2}(x)}x\left( 6+7\frac{t^{2/3}}{t_{today}^{2/3}}-9\frac{t^{4/3}}{%
t_{today}^{4/3}}\right) +\frac{3\eta _{today}^3}{2000}\sqrt{\frac{10\gamma }{%
3a^3}}\frac{\xi ^{^{\prime }2}(x)}x-  \nonumber \\
&&\left. -\frac{\eta _{today}^4}{2400a^2}\xi ^{^{\prime \prime }2}(x)\frac{%
t^{5/3}}{t_{today}^{5/3}}+\frac{\eta _{today}^4}{2400a^2}\xi ^{^{\prime
\prime }2}(0)+\frac{\eta _{today}^4}{1000a^2}\xi ^{^{\prime \prime }2}(x)%
\frac t{t_{today}}-\frac{\eta _{today}^4}{1000a^2}\xi ^{^{\prime \prime
}2}(0)\right\}  \nonumber
\end{eqnarray}
and here again, $N(x)=F(x)$. Observe that the new corrections are obviously
non linear in the perturbations.

The derivation of these formulae is valid irrespective of the details of the
primordial spectrum $\xi (x)$, and so this method may be used to work out in
a simple fashion the predictions of several scenarios of primordial
fluctuation generation.

\section{The view from Earth}

Due to the spherical symmetry of the Tolman Universe, it is not possible to
study anisotropic effects using radial geodesics, i.e. an observer located
in the center of the Universe, will detect no anisotropies. To study the
presence of anisotropies in a physical system using the Tolman solution to
Einstein equations we have to move the observer away from the center of
symmetry and solve for it the null geodesics equation. In this section we
solve the mentioned equation to both leading and next to leading order, and
find expressions to analyze the anisotropies in the CMBR due to the
different redshifts of photons, travelling from the decoupling surface to
the present.

Due to the spherical symmetry of the Tolman Universe we do not loose
generality if we locate the observer at $\theta =0$. In this case the
Hamilton Jacobi equation for the characteristic function reads 
\begin{equation}
-\left( \frac{\partial W}{\partial t}\right) ^2+\frac{f^2(x)}{R^{\prime
2}(x,t)}\left( \frac{\partial W}{\partial x}\right) ^2+\frac 1{R^2(x,t)}%
\left( \frac{\partial W}{\partial \theta }\right) ^2=-E  \label{genHJeq}
\end{equation}
where we used as the principal function $S=W+\lambda E$. We separate
variables by writing $W=T(t,p,E)+L(x,p,\alpha )+M(\alpha ,\theta )$.

\subsection{Leading Order Approximation}

We rewrite Eq. (\ref{genHJeq}) as 
\begin{equation}
-\left( \frac{\partial W}{\partial t}\right) ^2+\frac 1{v_0^2t^{4/3}}\left( 
\frac{\partial W}{\partial x}\right) ^2+\frac 1{v_0^2x^2t^{4/3}}\left( \frac{%
\partial W}{\partial \theta }\right) ^2=-E  \label{loHJeq}
\end{equation}
with $v_0^2=9t_{today}^{2/3}/x_0^2$ and where we have used (\ref
{fgran-approx}) to evaluate $R(x,t)$ and $R^{^{\prime }}(x,t)$. In this case
we have 
\begin{equation}
\frac{dM}{d\theta }=v_0\alpha \rightarrow M=v_0\alpha \theta  \label{M}
\end{equation}
\begin{equation}
\frac{dL}{dx}=P_r=v_0\sqrt{p^2-\frac{\alpha ^2}{x^2}}\rightarrow L=v_0\left[ 
\sqrt{p^2x^2-\alpha ^2}-\alpha \arccos \left( \frac \alpha {px}\right)
\right]  \label{L}
\end{equation}
\begin{equation}
\frac{dT}{dt}=\sqrt{E+\frac{p^2}{t^{4/3}}}  \label{dT/dt}
\end{equation}
As we are interested in a neighborhood of $E=0$, we can expand 
\begin{equation}
\frac{dT}{dt}\simeq \frac p{t^{2/3}}+\frac 12E\frac{t^{2/3}}p-\frac 18E^2%
\frac{t^2}{p^3}+\cdots  \label{dT/dt-devel}
\end{equation}
which can be integrated to give 
\begin{equation}
T\simeq 3pt^{1/3}+\frac 3{10}E\frac{t^{5/3}}p-\frac 1{24}E^2\frac{t^3}{p^3}%
+\cdots  \label{T-devel}
\end{equation}
If we define new canonical variables $\tau $, $\tilde \theta $ and $\rho $
conjugated to $E$, $\alpha $ and $p$ respectively, we have from the
characteristic function $W$ that 
\begin{equation}
\tau =\frac{\partial W}{\partial E}=\frac 3{10}\frac{t^{5/3}}p-\frac 1{12}E%
\frac{t^3}{p^3}+\cdots  \label{tau-gen}
\end{equation}
\begin{equation}
\rho =\frac{\partial W}{\partial p}=\frac{v_0}p\sqrt{p^2x^2-\alpha ^2}%
+3t^{1/3}-\frac 3{10}E\frac{t^{5/3}}{p^2}+\cdots  \label{rho-gen}
\end{equation}
\begin{equation}
\chi =\frac{\partial W}{\partial \alpha }=v_0\left[ \theta -\arccos \left( 
\frac \alpha {px}\right) \right]  \label{xi-gen}
\end{equation}
>From (\ref{rho-gen}) and recalling that $F^{1/3}(x)\simeq
(12t_{today})^{1/3}x/x_0$ we get 
\begin{equation}
\frac{x_0}{(12t_{today})^{1/3}}F^{1/3}(x)=\frac 1p\sqrt{\frac{p^2}{\nu _0^2}%
\left( \rho -3t^{1/3}+\frac 3{10}E\frac{t^{5/3}}{p^2}-\cdots \right)
^2+\alpha ^2}  \label{x-gen}
\end{equation}
from where we will read the expression for $F^{1/3}(x)$.

We can write the old momentum $P_t$ in terms of the new variables as 
\begin{equation}
P_t=\frac{\partial W}{\partial t}=\sqrt{E+\frac{p^2}{t^{4/3}}}\simeq \frac p{%
t^{2/3}}+\frac 12E\frac{t^{2/3}}p-\frac 18E^2\frac{t^2}{p^3}+\cdots
\label{Pt-NC-gen}
\end{equation}

The new Hamiltonian is $H=-E$. The equations of motion are 
\begin{eqnarray}
\dot \tau &=&\partial H/\partial E=-1  \label{Hameq-gen} \\
\dot E &=&\dot p=\dot \rho =\dot \alpha =\dot \chi =0  \nonumber
\end{eqnarray}
where a dot means $d/d\lambda $. As $p=const$ we can choose it as 
\begin{equation}
p=t_{today}^{2/3}  \label{p-gen}
\end{equation}
so that 
\begin{equation}
P_t\mid _{E=0}=(1+z)=\left( \frac{t_{today}}t\right) ^{2/3}  \label{Pt-LO}
\end{equation}
therefore when $t=t_{today}$ we have $z=0$. In order to determine $\alpha $
we need to specify the position of the observer, $x_T$ and the angle of
observation, $\varphi $. We calculate the observer's position by adjusting
the velocity of the Earth to the value determined by the dipolar
contribution to the background radiation anisotropy, i.e. 
\begin{equation}
\dot R=2t_{today}^{1/3}\frac x{x_0}t^{-1/3}\mid _{t=today}=1.666\times
10^{-3}\rightarrow x_T=0.833\times 10^{-4}x_0  \label{xT-det}
\end{equation}
The angle subtended by a null ray that reaches $x_T$ at $t_{today}$ is
defined through (see Fig. {\bf 1}) 
\begin{equation}
v_0\cos (\pi -\varphi )=\frac{P_r}p  \label{ang-obs}
\end{equation}
from where we obtain 
\begin{equation}
\alpha =px_T\sin \varphi  \label{alpha}
\end{equation}
As $\rho $ is a constant of the motion, we fix its value by replacing in eq.
(\ref{rho-gen}) $x=x_T$, $t=t_{today}$, i.e. we choose 
\begin{equation}
\rho =3t_{today}\left[ 1+\frac x{x_0}\cos \varphi \right]  \label{rho-LO-fix}
\end{equation}
Therefore the equation for the geodesics reads 
\begin{equation}
x=x_0\sqrt{\left( 1-\frac{t^{1/3}}{t_{today}^{1/3}}\right) ^2+2\left( 1-%
\frac{t^{1/3}}{t_{today}^{1/3}}\right) \frac{x_T}{x_0}\cos \varphi +\frac{%
x_T^2}{x_0^2}}  \label{x-gen-LO}
\end{equation}
and the redshift is given by 
\begin{equation}
1+z=\frac{t_{today}^2}{t^{2/3}}  \label{rdsh-noncen}
\end{equation}

\subsection{Next to Leading Order Approximation}

In this case the Hamilton-Jacobi equation for the characteristic function
reads 
\begin{equation}
-\left( \frac{\partial W}{\partial t}\right) ^2+\frac 1{v_0^2t^{4/3}}%
[1-v_1(x)t^{2/3}]\left( \frac{\partial W}{\partial x}\right) ^2+\frac 1{%
v_0^2t^{4/3}}[1-w_1(x)t^{2/3}]\left( \frac{\partial W}{\partial \theta }%
\right) ^2=-E  \label{NTLOHJeq}
\end{equation}
where 
\begin{eqnarray}
v_1 &=&\frac 1{10}\frac{x_0^2}{t_{today}^{2/3}}\left[ \frac{f^2(x)-1}x%
\right] ^{\prime }=-\frac{\eta ^2}{10at_{today}^{2/3}}\xi ^{^{\prime \prime
}}(x)  \label{v1(x)-w1(x)} \\
w_1 &=&\frac 1{10}\frac{x_0^2}{t_{today}^{2/3}}\left[ \frac{f^2(x)-1}{x^2}%
\right] =-\frac{\eta ^2}{10at_{today}^{2/3}}\frac{\xi ^{\prime }(x)}x 
\nonumber
\end{eqnarray}
(cfr. Appendix {\bf A})

Now, $E$, $p$ and $\alpha $ are no longer constants of motion because in
terms of them 
\begin{equation}
H=-E-p^2\frac{v_1(x)}{t^{2/3}}-\frac{\alpha ^2}{x^2t^{2/3}}\left[
w_1(x)-v_1(x)\right]  \label{H-NC-O1}
\end{equation}
We therefore perform further canonical transformation to variables $E_1$, $%
p_1$, $\alpha _1$, $\rho _1$, $\chi _1$, $\tau _1$ so that $H=-E_1$. The
generating function is now 
\begin{equation}
W_1=E_1\tau +p_1\rho +\alpha _1\chi -p_1^2{\cal G}_1(\rho ,\chi ,\tau
,p_1,\alpha _1,E_1)-\alpha _1^2{\cal F}_1(\rho ,\chi ,\tau ,p_1,\alpha
_1,E_1)  \label{W1-NC}
\end{equation}
The old momentum $E$ becomes 
\begin{equation}
E=\frac{\partial W_1}{\partial \tau }=E_1-p_1^2\frac{\partial {\cal G}_1}{%
\partial \tau }-\alpha _1^2\frac{\partial {\cal F}_1}{\partial \tau }
\label{E-NC-def}
\end{equation}
so to obtain $H=-E_1$ we demand 
\begin{equation}
\frac{\partial {\cal G}_1}{\partial \tau }=\frac{v_1(x)}{t^{2/3}}
\label{dG1/dtau-NC}
\end{equation}
\begin{equation}
\frac{\partial {\cal F}_1}{\partial \tau }=\frac 1{x^2t^{2/3}}\left[
w_1(x)-v_1(x)\right]  \label{dF1/dtau-NC}
\end{equation}
We can write formally 
\begin{equation}
{\cal G}_1=\int\limits_{\tau _{today}}^\tau d\tau ^{\prime }\frac{v_1[x(\rho
,\chi ,\tau ^{\prime },p_1,\alpha _1,E_1)]}{t^{2/3}(E_1,p_1,\tau ^{\prime })}
\label{G1-NC-form.}
\end{equation}
\begin{equation}
{\cal F}_1=\int\limits_{\tau _{today}}^\tau d\tau ^{\prime }\frac{w_1[x(\rho
,\chi ,\tau ^{\prime },p_1,\alpha _1,E_1)]-v_1[x(\rho ,\chi ,\tau ^{\prime
},p_1,\alpha _1,E_1)]}{x^2(\rho ,\chi ,\tau ^{\prime },p_1,\alpha
_1,E_1)\;t^{2/3}(E_1,p_1,\tau ^{\prime })}  \label{F1-NC-form.}
\end{equation}
In terms of the zeroth order variables we can write (holding fixed $E_1$, $%
p_1$) 
\begin{equation}
\tau =\frac{dT}{dE}=\frac d{dE}\left[ \int\limits_{t_{today}}^tdt\frac p{%
t^{2/3}}\sqrt{1+E\frac{t^{4/3}}{p^2}}\right] \rightarrow d\tau =\frac 1{2p}%
\frac{t^{2/3}dt}{\sqrt{1+E\frac{t^{4/3}}{p^2}}}  \label{dtau(t)-NC}
\end{equation}
so 
\begin{equation}
{\cal G}_1=\frac 1{2p_1}\int\limits_{\tau _{today}}^\tau \frac{dt}{\sqrt{1+E%
\frac{t^{4/3}}{p^2}}}v_1[x(\rho ,\chi ,\tau ,p_1,\alpha _1,E_1)]
\label{G1(t)-NC-O1}
\end{equation}
\begin{equation}
{\cal F}_1=\frac 1{2p_1}\int\limits_{\tau _{today}}^\tau \frac{dt}{\sqrt{1+E%
\frac{t^{4/3}}{p^2}}}\frac{w_1[x(\rho ,\chi ,\tau ,p_1,\alpha
_1,E_1)]-v_1[x(\rho ,\chi ,\tau ,p_1,\alpha _1,E_1)]}{x^2(\rho ,\chi ,\tau
,p_1,\alpha _1,E_1)}  \label{F1(t)-NC-O1}
\end{equation}
where $t=t(E_1,p_1,\tau )$. Let us now compute the corrections to the
several quantities introduced so far.

\subsubsection{Corrections to $E$}

The value of $E$ valid to first order follows immediately from eqs. (\ref
{E-NC-def}),(\ref{dG1/dtau-NC}) and (\ref{dF1/dtau-NC}) plus the Hamiltonian
constraint $E_1=0$, leading to 
\begin{equation}
E=-p_1^2\frac{v_1(x)}{t^{2/3}}-\alpha _1^2\frac{w_1(x)-v_1(x)}{x^2t^{2/3}}
\label{E-NC-O1}
\end{equation}
where $t=t(E_1=0,p_1,\tau l1)$, $x=x(E_1=0,p_1,\alpha _1,\rho _1,\tau _1)$.

\subsubsection{Corrections to $p$}

The value of the old momentum $p$ follows from 
\begin{equation}
p=\frac{\partial W_1}{\partial \rho }=p_1-p_1^2\frac{\partial {\cal G}_1}{%
\partial \rho }-\alpha _1^2\frac{\partial {\cal F}_1}{\partial \rho }
\label{p-NC-def}
\end{equation}
Eqs. (\ref{G1(t)-NC-O1}) and (\ref{F1(t)-NC-O1}) lead to 
\begin{equation}
\frac{\partial {\cal G}_1}{\partial \rho }=\frac 1{2p_1}\int%
\limits_{t_{today}}^t\frac{dt}{\sqrt{1+E_1\frac{t^{\prime 4/3}}{p^2}}}%
v_1^{^{\prime }}(x)\frac{dx}{d\rho }  \label{dG1/drho-NC}
\end{equation}
\begin{equation}
\frac{\partial {\cal F}_1}{\partial \rho }=\frac 1{2p_1}\int%
\limits_{t_{today}}^t\frac{dt}{\sqrt{1+E_1\frac{t^{\prime 4/3}}{p^2}}}\left[ 
\frac{w_1(x)-v_1(x)}{x^2}\right] ^{^{\prime }}\frac{dx}{d\rho }
\label{dF1/drho-NC}
\end{equation}
Setting $E_1=0$, we have from the zeroth order 
\begin{equation}
\frac{\partial x}{\partial \rho }=\frac{\rho -3t^{1/3}}{v_0^2\sqrt{\frac{%
(\rho -3t^{1/3})^2}{v_0^2}+\frac{\alpha ^2}{p^2}}}=\frac{\rho -3t^{1/3}}{%
v_0^2x}=\frac{\sqrt{p_1^2x^2-\alpha _1^2}}{v_0p_1x}  \label{dx/drho-NC}
\end{equation}
and consequently 
\begin{eqnarray}
p &=&p_1-\frac{p_1}2\int\limits_{t_{today}}^t\frac{dt}{\sqrt{1+E_1\frac{%
t^{\prime 4/3}}{p_1^2}}}v_1^{^{\prime }}(x)\frac{\sqrt{p_1^2x^2-\alpha _1^2}%
}{v_0p_1x}-  \label{p(t)-NC-O1} \\
&&-\frac{\alpha _1^2}{2p_1}\int\limits_{t_{today}}^t\frac{dt}{\sqrt{1+E_1%
\frac{t^{\prime 4/3}}{p_1^2}}}\left[ \frac{w_1(x)-v_1(x)}{x^2}\right]
^{^{\prime }}\frac{\sqrt{p_1^2x^2-\alpha _1^2}}{v_0p_1x}  \nonumber
\end{eqnarray}

\subsubsection{Corrections to $\alpha $}

The value of the old momentum $\alpha $ follows from 
\begin{equation}
\alpha =\frac{\partial W_1}{\partial \chi }=\alpha _1  \label{alpha-NC-O1}
\end{equation}
a result that could be guessed due to the fact that the perturbations do not
depend on the angles.

\subsubsection{Corrections to $\rho $ and $\tau $}

The new coordinates are 
\begin{equation}
\rho _1=\frac{\partial W_1}{\partial p_1}=\rho -2p_1G_1-p_1^2\frac{\partial 
{\cal G}_1}{\partial p_1}-\alpha _1^2\frac{\partial {\cal F}_1}{\partial p_1}
\label{rho1-NC-def}
\end{equation}
At $E_1=0$ we have from eqs. (\ref{G1(t)-NC-O1}) and (\ref{F1(t)-NC-O1})
that 
\begin{equation}
\frac{\partial {\cal G}_1}{\partial p_1}=-\frac 1{2p_1^2}\int\limits_{\tau
_{today}}^\tau dt\,v_1(x)+\frac 1{2p_1}v_1(x)\frac{\partial t}{\partial p_1}+%
\frac 1{2p_1}\int\limits_{\tau _{today}}^\tau dt\,v_1^{^{\prime }}(x)\frac{%
\partial x}{\partial p_1}  \label{dG1/dp1-NC}
\end{equation}
\begin{eqnarray}
\frac{\partial {\cal F}_1}{\partial p_1} &=&-\frac 1{2p_1^2}%
\int\limits_{\tau _{today}}^\tau dt^{\prime }\left( \frac{w_1(x)-v_1(x)}{x^2}%
\right) +\frac 1{2p_1}\left( \frac{w_1(x)-v_1(x)}{x^2}\right) \frac{\partial
t}{\partial p_1}+  \label{df1/dp1-NC} \\
&&+\frac 1{2p_1}\int\limits_{\tau _{today}}^\tau dt^{\prime }\frac d{dx}%
\left( \frac{w_1(x)-v_1(x)}{x^2}\right) \frac{\partial x}{\partial p_1} 
\nonumber
\end{eqnarray}
>From the zeroth order, if $E=0$, we have that $t=(10p\tau /3)^{3/5}$, so
that $\partial t/\partial p=3t/5p_1$. Also from this order 
\begin{equation}
x=\sqrt{\frac{\left( \rho -3t^{1/3}\right) ^2}{v_0^2}+\frac{\alpha ^2}{p^2}}%
\rightarrow \frac{\partial x}{\partial p}=\frac{-\alpha ^2}{p^3\sqrt{\frac{%
\left( \rho -3t^{1/3}\right) ^2}{v_0^2}+\frac{\alpha ^2}{p^2}}}=\frac{%
-\alpha _1^2}{p_1^3x}  \label{dx/dp-NC}
\end{equation}
Therefore we have 
\begin{eqnarray}
\rho &=&\rho _1+\frac 12\int\limits_{\tau _{today}}^\tau dt\,v_1(x)+\frac 3{%
10}v_1(x)t+\frac 3{10}\frac{\alpha _1^2}{p_1^2}\left( \frac{w_1(x)-v_1(x)}{%
x^2}\right) t-  \nonumber \\
&&-\frac{\alpha _1^2}{2p_1^2}\int\limits_{t_{today}}^tdt\frac{w_1(x)-v_1(x)}{%
x^2}-\frac{\alpha _1^2}{2p_1^2}\int\limits_{\tau _{today}}^\tau dt\frac{%
v_1^{^{\prime }}(x)}x-  \label{rho-NC-O1} \\
&&-\frac{\alpha _1^4}{2p_1^4}\int\limits_{\tau _{today}}^\tau dt\left( \frac{%
w_1^{^{\prime }}(x)-v_1^{^{\prime }}(x)}{x^3}\right) +\frac{\alpha _1^4}{%
p_1^4}\int\limits_{t_{today}}^tdt\frac{w_1(x)-v_1(x)}{x^4}  \nonumber
\end{eqnarray}

In order to compute the integrals, it is convenient to parametrize the null
geodesic with the coordinate $x$, from eq. (\ref{x-gen-LO}) we get 
\begin{equation}
dt=-\left\{ 1+\frac{v_0}{3pt_{today}}\left[ \sqrt{p_1^2x_T^2-\alpha _1^2}-%
\sqrt{p_1^2x^2-\alpha _1^2}\right] \right\} ^2\frac{v_0t_{today}^{2/3}p_1x}{%
\sqrt{p_1^2x^2-\alpha _1^2}}dx  \label{dt(x)-NC}
\end{equation}
Therefore the corrections computed above read 
\begin{equation}
p=p_1+\frac{p_1}2t_{today}^{2/3}v_1(x)+\frac{\alpha _1^2}{2p_1}%
t_{today}^{2/3}\frac{\left( w_1(x)-v_1(x)\right) }{x^2}  \label{p-NC-final}
\end{equation}
\begin{eqnarray}
\rho &=&\rho _1-t_{today}^{2/3}p_1v_0\int\limits_0^xdx\,\frac{\,x}{\sqrt{%
p^2x^2-\alpha ^2}}v_1(x)+\frac{3t}{10}v_1(x)+\frac{3\alpha _1^2}{10p_1^2}t%
\frac{w_1(x)-v_1(x)}{x^2}+  \label{rho-NC-final} \\
&&\ \ +\frac{\alpha _1^2}{2p_1}t_{today}^{2/3}v_0\int\limits_0^x\frac{dx}{%
\sqrt{p^2x^2-\alpha ^2}}v_1^{\prime }(x)+\frac{\alpha _1^2}{p_1}%
t_{today}^{2/3}v_0\int\limits_0^x\frac{dx}{\sqrt{p^2x^2-\alpha ^2}}\left( 
\frac{w_1(x)-v_1(x)}x\right) +  \nonumber \\
&&\ \ \frac{\alpha _1^4}{2p_1^3}t_{today}^{2/3}v_0\int\limits_0^x\frac{dx}{%
\sqrt{p^2x^2-\alpha ^2}}\frac{w_1^{\prime }(x)-v_1^{\prime }(x)}{x^2}-\frac{%
\alpha _1^4}{p_1^3}t_{today}^{2/3}v_0\int\limits_0^x\frac{dx}{\sqrt{%
p^2x^2-\alpha ^2}}\frac{w_1(x)-v_1(x)}{x^3}  \nonumber
\end{eqnarray}
where $x$ is to be read from the leading order and where $v_1(x)$ and $%
w_1(x) $ are given by eq. (\ref{v1(x)-w1(x)}).

Now we are ready to evaluate the next to leading order corrections to the
redshift and $F^{1/3}(x)$. For the redshift we have to replace eqs. (\ref
{E-NC-O1}) and (\ref{p-NC-final}) into eq. (\ref{Pt-NC-gen}), keeping only
the leading and next to leading order terms. We get 
\begin{equation}
P_t=\frac{p_1}{t^{2/3}}\left\{ 1-\frac{\eta _{today}^2}{20a^2}\xi ^{^{\prime
\prime }}(x)\left( 1-\frac{t^{2/3}}{t_{today}^{2/3}}\right) -\frac{\eta
_{today}^2}{20a^2}x_T^2\sin ^2\varphi \left[ \frac{\xi ^{^{\prime }}(x)}{x^3}%
-\frac{\xi ^{^{\prime \prime }}(x)}{x^2}\right] \left( 1-\frac{t^{2/3}}{%
t_{today}^{2/3}}\right) \right\}  \label{Pt-NC-O1}
\end{equation}
where we have used $\alpha _1=p_1x_T\sin \varphi $.

In order to evaluate $F^{1/3}(x)$ we recall expression (\ref{x-gen}) and
replace in it eqs. (\ref{rho-NC-final}) and (\ref{E-NC-O1}) obtaining 
\begin{equation}
F^{1/3}(x)=\left( 12t_{today}\right) ^{1/3}\sqrt{\left[ 1+\frac{x_T}{x_0}%
\cos \varphi -\frac{t^{1/3}}{t_{today}^{1/3}}+I(x)\right] ^2+\frac{x_T^2}{%
x_0^2}\sin ^2\varphi }  \label{F(x)-NC-O1}
\end{equation}
where 
\begin{eqnarray}
I(x) &=&\frac{\eta _{today}}{20}\sqrt{\frac{10\gamma }{3a}}%
\int\limits_{x_T}^x\frac{x\,dx}{\sqrt{x^2-x_T^2\sin ^2\varphi }}\xi
^{^{\prime \prime }}(x)- \\
&&-\frac{\eta _{today}}{20}\sqrt{\frac{10\gamma }{3a}}x_T^2\sin ^2\varphi
\int\limits_{x_T}^x\frac{dx}{\sqrt{x^2-x_T^2\sin ^2\varphi }}\xi ^{^{\prime
\prime \prime }}(x)-  \nonumber \\
&&-\frac{\eta _{today}}{20}\sqrt{\frac{10\gamma }{3a}}x_T^2\sin ^2\varphi
\int\limits_{x_T}^x\frac{dx}{\sqrt{x^2-x_T^2\sin ^2\varphi }}\left[ \frac{%
\xi ^{^{\prime }}}{x^2}-\frac{\xi ^{^{\prime \prime }}(x)}x\right] +
\label{I(x)} \\
&&+\frac{\eta _{today}}{20}\sqrt{\frac{10\gamma }{3a}}x_T^4\sin ^4\varphi
\int\limits_{x_T}^x\frac{dx}{\sqrt{x^2-x_T^2\sin ^2\varphi }}\left[ 3\frac{%
\xi ^{^{\prime }}}{x^4}-3\frac{\xi ^{^{\prime \prime }}(x)}{x^3}+\frac{\xi
^{^{\prime \prime \prime }}}{x^2}\right]  \nonumber
\end{eqnarray}
In both equations (\ref{Pt-NC-O1}) and (\ref{F(x)-NC-O1}) we have imposed as
boundary conditions $P_t=1$ and $F^{1/3}(x)=(12t_{today})^{1/3}x_T/x_0$ at $%
x=x_T$, $t=t_{today}$.

As earlier, our formulae are valid for any choice of $\xi (x)$.

\section{An Example: Scale - Invariant Primordial Perturbations}

In order to apply our developments to a concrete case, we must now specified 
$\zeta _k$ in Eq. (\ref{C2-superp}). In the absence of a specific model for
the spectrum of a primeval fluctuations, it is often assumed that, at
horizon crossing $\left| \zeta _k\right| =\beta k^{(n-4)/2}$. On a $t=const.$
surface the corresponding expression is 
\begin{equation}
\left| \zeta _k\right| =\beta k^{(n-3)/2}  \label{zetak-t=cte}
\end{equation}
where $n$ is known as the spectral index. This index is determined by the
equation of state $\omega =p/\rho $ at $50-60$ e-folds before the end of
Inflation, and its expression for scalar perturbations is $n=1-3(1+\omega
)+d\ln (1+\omega )/d\ln k$, where $k$ is the wavenumber \cite{Steinhardt}.
Strict exponential (de Sitter) expansion corresponds to $n=1$, the so-called
Harrison - Zel'dovich spectrum. But in any realistic inflationary model, the
expansion rate must slow down near the end of Inflation in order to return
to FRW expansion, and in this case $n\neq 1$ is expected\cite{Steinhardt}.
The specification of the spectrum is concluded by also fixing the phases of
the amplitudes for the different modes, which in inflationary models are
usually random Gaussian variables; for simplicity, we shall disregard this,
assuming $\left| \zeta _k\right| =\zeta _k$ instead. Of course, applying our
formulae to the more general case of nontrivial phases involves no
difficulties of principle.

With the form (\ref{zetak-t=cte}), we have 
\begin{equation}
\gamma =\beta k_0^{(n+3)/2}  \label{gamma-t=cte}
\end{equation}
and 
\begin{eqnarray}
\xi (x) &=&\frac 1x\int\limits_0^1\frac{du}uu^{(n+1)/2}\sin ux
\label{Xi(x)-t=cte} \\
\ &=&-\frac i{(n+1)x}\left[ _1F_1\left( \frac{n+1}2;\frac{n+3}2;ix\right)
-\,_1F_1\left( \frac{n+1}2;\frac{n+3}2;-ix\right) \right]  \nonumber
\end{eqnarray}
where $_1F_1\left( \frac{n+1}2;\frac{n+3}2;\pm ix\right) $ are the
degenerate hypergeometric function \cite{Gradshteyn}. From eqs. (\ref
{fgran(x)}) and (\ref{fchic(x)}) we see that in general $F(x)$ and $f^2(x)$
will be oscillatory functions of $x$. The effect of the perturbations at the
beginning of the matter dominated epoch is a mass redistribution, and the
creation of negative curvature regions ($f^2(x)>1$), that will expand
forever, and of positive curvature ones ($f^2(x)$ $<1$), that will
eventually collapse. This distribution of matter will evolve non linearly
from its initial state at $t_0$, its evolution being described by the Tolman
equations. We normalize the spectrum by considering the amplitude of the
anisotropy of the cosmic background radiation corresponding to a scale of
the size of the horizon at decoupling ($\lambda _d\simeq 0.13\,Mpc$) \cite
{partridge}. 
\begin{equation}
k_d^3\zeta _{k_d}=\gamma _nk_d^{(n+3)/2}\equiv \gamma \left( \frac{k_d}{k_0}%
\right) ^{(n+3)/2}\simeq 6.6\times 10^{-5}  \label{k3zeta(x)-norm}
\end{equation}
where we used $k_d^3\zeta _{k_d}\simeq 3\Delta T/T$. By virtue of the linear
relationship between the angles that the scales subtend today and their
sizes, we have $k_d/k_0\sim \theta _0/\theta _d$, where $\theta _d$ is the
angle subtended today by a scale of the size of the horizon at decoupling, $%
\theta _d\sim 1.4^{\circ }-2^{\circ }$, and $\theta _0$ is the angle
subtended today by the wavelength corresponding to $k_0$, $\theta _0\sim
45^{\prime }-1^{\circ }$. For concreteness we choose $\theta _0\sim
48^{\prime }$, $\theta _d\sim 1.4^{\circ }$.

With the form (\ref{Xi(x),n=1}) for $\xi (x)$ we have $a=1/12$. If we take $%
k_0\simeq 918hMpc^{-1}$ (which corresponds to a wavelength about six times
bigger than the horizon size at equilibrium), $t_0\simeq 5.65\times
10^{-4}h^{-1}Mpc$, $t_{today}\simeq 1998h^{-1}Mpc$ and $\gamma \simeq
2.02125\times 10^{-4}$, we get $\eta \simeq 1.78$ (cfr. eq. (\ref{etadef})).

The model becomes particularly simple when $n=1$, whereby eq. (\ref
{Xi(x)-t=cte}) becomes 
\begin{equation}
\xi (x)=\frac 1{x^2}\left[ 1-\cos x\right]  \label{Xi(x),n=1}
\end{equation}
This function is to be used in expressions found in the previous sections.

\subsection{Density Contrast, Number Counts and Luminosity Distance}

In Fig. {\bf 2} we have plotted the density contrast $\Delta \delta /\bar 
\delta $ vs $z$ for the computed second order. As the matching conditions
tell us nothing about the sign of $f^2(x)-1$, we chose it so that the
perturbation evolved to form a void. We see that the profile is of non
compensated kind. After the non linear evolution, the density contrast in
the wall is $\Delta \delta /\bar \delta \simeq 0.06$ and in the center of
the void $\Delta \delta /\bar \delta \simeq -0.44$. If we define the radius
of the void as the distance between its center and the point where $\Delta
\delta /\bar \delta =0$ (which corresponds to $z\simeq 0.024$), we get $%
R\simeq 72h^{-1}Mpc$, taking into account the results of Fig. {\bf 4}.

In Fig. {\bf 3} we have plotted $\ln N(x)$ vs $\ln z$. Dotted line
corresponds to an unperturbed Universe, full line to one with second order
perturbation. We see that the number count increases with $z$ as we move
over the increasing density zone. For $z\simeq 0.05$ the number count begins
to equal that for a FRW Universe.

In order to analyze the departures from the linear Hubble law due to the
presence of the perturbations, we have plotted in Fig. {\bf 4} the $z$ vs $%
d_\ell $ relationship, for a flat FRW Universe (dotted line) and for Tolman
Universe (full line). Due to the small redshifts considered, the slope of
the curves can be considered a measure of the Hubble constant, which in our
units is $H_0=2$ (the true value of the Hubble constant being $%
H_0=2/3t_{today}$). We see that the slope of the curve corresponding to the
perturbed Universe is smaller than the one corresponding to FRW, up to $%
z\simeq 0.013$. From there on the slope starts to increase to finally
coincide with the unperturbed one as expected.

\subsection{Anisotropies in the CMBR}

In our Tolman Universe, we find two sources of CMBR anisotropies, one of
which is due to the fluctuations in the density, $\Delta T/T\propto \Delta
\delta /\bar \delta $ ($\Delta \delta /\bar \delta $ given by eq.(\ref
{dencon-c-O1}) evaluated using eq. (\ref{x-gen-LO})). The origin of the
other is that being the Earth away from the center of symmetry, the light
that was emitted from different points of last scattering surface will reach
the observer with different redshifts, depending on the angular position of
the source point, i.e. we have a contribution to the CMBR anisotropies of
the form $z(\theta )-\bar \zeta $, where $\bar \zeta $ is the mean redshift
of the decoupling surface. We find that in our case the last contribution
outweights the first one by a factor of $10^7$: we calculated $\Delta \delta
/\bar \delta $ and $z(\theta )-\bar \zeta $ and after substracting the
dipolar contribution due to the movement of the Earth ($\Delta T/T\approx
10^{-3}$) we were left with a quadrupolar contribution of $\Delta T/T\approx
10^{-13}$ and $\Delta T/T\approx 10^{-6}$, respectively, for big angular
scales. The fact that the contribution from the density fluctuations is so
small is not surprising in view of the small cut-off $k_0$.

\section{Discussion}

In this paper we have developed an analytical method to study the non linear
evolution of adiabatic perturbations in a matter dominated Universe. We
build the initial profile of the fluctuation on a $t=const.$ surface (that
corresponds to the beginning of the matter dominated era, when structure
formation begins) as a superposition of all those modes whose wavenumber is
smaller than a certain cut-off $k_0$. This cut-off may correspond to the
mode that survived the Meszaros effect \cite{meszaros}, Landau damping or
free streaming \cite{Peebles93}. 

>From this point on we can follow the non linear evolution of the
perturbations by means of the exact solution to Einstein equation for a
pressureless, spherically symmetric Universe, namely the Tolman Universe.
Since this solution is isotropic with respect to one point, the Tolman
solution can be applied only to analyze the formation of a structure (for
example a void or cluster) with the proper symmetry. This restriction in the
applicability of the Tolman solution to the study of structure evolution is
compensated by the advantages of having an exact solution at our disposal.

Formally, the basis of all calculations for observable quantities lays in
the resolution of the equation for the null geodesics, which in general is
not trivial and is carried out by numerical methods. In this work we
presented a perturbative method to do this calculation analytically. We put
our calculations in a physical framework by the way in which we build the
perturbations. We carried out the calculations only to second order, which
means a mild non linear evolution, but the extension to higher orders is
straightforward.

In the last section we obtained a glimpse of the results to be obtained by
assuming a scale free spectrum of initial perturbations. In view of the
simplicity of the calculations involved, we find that our results give a
remarkable approximation to the structure of some of the largest known
voids, such as B\"otes'.Quantitatively, our results are accurate only up to
an order of magnitude (in fact, the B\"otes void is surrounded by a wall
whose density contrast is $\Delta \delta /\bar \delta \simeq 4$, and its
radius is of about $30Mpc$ \cite{deLapparent}). The difference, of course,
could be reduced by a better choice of the cutoff wavelength and, most
importantly, by carrying the computation to higher orders.

With regard to future work, the most important feature of the method we
propose, besides its analytical and relatively simple character, is that it
may be applied for any form of the primordial spectrum. Thus it becomes only
a matter of plugging in one's favorite theory of primordial fluctuation
generation (as reflected in the particular form of $\zeta _k$ (Eq. (\ref
{C2-superp})) to easily obtain (rough) testable predictions of that model.
While we have used here a (unrealistic) scale invariant spectrum for
demonstration purposes, the calculation is equally simple with red, blue, or
more sophisticated alternatives.

We therefore believe the methods we describe in this paper will be an useful
tool in the delicate task of sorting between the manyfold fluctuation
generation scenarios now available.

\section{Acknowledgments}

It is a pleasure to thank N. Deruelle, D. Harari and M. Zaldarriaga for
comments on earlier versions of this work, and C. El Hasi, A. Gangui, and S.
Gonorazky for their help on carrying it to completion. This work has been
partially supported by Universidad de Buenos Aires, CONICET and Fundaci\'on
Antorchas, and by the Commission of the European Communities under Contract
CI1*-CJ94-0004.

\section{Appendix A: Expansion of the radial Tolman function}

\begin{equation}
R(x,t)=\frac{12^{2/3}}4F^{1/3}(x)t^{2/3}\left\{ 1+\frac{12^{2/3}}{20}\frac{%
\left[ f^2(x)-1\right] }{F^{2/3}(x)}t^{2/3}-\frac 3{2800}12^{4/3}\frac{%
\left[ f^2(x)-1\right] ^2}{F^{4/3}(x)}t^{4/3}+\cdots \right\}  \label{R(x,t)}
\end{equation}

\begin{eqnarray}
\frac 1{R^2(x,t)} &=&\frac{4^2}{12^{4/3}}\frac 1{F^{2/3}(x)t^{4/3}}\left\{ 1-%
\frac{12^{2/3}}{10}\frac{[f^2(x)-1]}{F^{2/3}(x)}t^{2/3}+\frac 3{1400}12^{4/3}%
\frac{[f^2(x)-1]^2}{F^{4/3}(x)}t^{4/3}+\right.  \label{rcuad} \\
&&\ \left. \ \ +\frac 3{400}12^{4/3}\frac{[f^2(x)-1]^2}{F^{4/3}(x)}%
t^{4/3}+\cdots \right\}  \nonumber
\end{eqnarray}
\begin{eqnarray}
\frac 1{R^{\prime 2}(x,t)} &=&\frac{4^2}{12^{4/3}}\frac 1{\left[
F^{1/3}(x)\right] ^{^{\prime }2}t^{4/3}}\left\{ 1-\frac 1{10}\frac{12^{2/3}}{%
\left[ F^{1/3}(x)\right] ^{^{\prime }}}\left[ \frac{f^2(x)-1}{F^{1/3}(x)}%
\right] ^{\prime }t^{2/3}+\right.  \nonumber \\
&&\ \ \ +\frac 3{400}\frac{12^{4/3}}{\left[ F^{1/3}(x)\right] ^{^{\prime }2}}%
\left[ \frac{f^2(x)-1}{F^{1/3}(x)}\right] ^{^{\prime }2}t^{4/3}+
\label{rprimcuad} \\
&&\left. +\frac 3{1400}\frac{12^{4/3}}{\left[ F^{1/3}(x)\right] ^{^{\prime }}%
}\left[ \frac{\left( f^2(x)-1\right) ^2}{F(x)}\right] ^{\prime
}t^{4/3}-\cdots \right\}  \nonumber
\end{eqnarray}

\section{Appendix B: Evaluation of $\Omega $}

We evaluate $\Omega $ today as follows. In a neighborhood of the origin, the
evolution of the Universe is nearly like a FRW one, therefore 
\begin{equation}
\Omega =1+\frac{{\cal K}}{H^2a^2(t)}=1+\frac{{\cal K}}{\dot a^2(t)}
\label{Omega(t)}
\end{equation}
where $H$ is the Hubble constant, $a(t)$ the expansion factor of the
Universe and ${\cal K}$ the curvature. We have to extract the expressions
for $\dot a(t)$ and ${\cal K}$ from the Tolman equations (\ref{fchic}), (\ref
{R-tol-open}) and (\ref{t-tol-open}). We define 
\begin{equation}
{\cal K}\equiv -{\frac 12}\frac{d^2f^2(x)}{dx^2}\mid _{x=0}
\label{K-Tol-def}
\end{equation}
from where we get ${\cal K}=(10/3)\gamma \xi ^{^{\prime \prime }}(0)\simeq
-5\times 10^{-5}$. To obtain $\dot a(t)$ we calculate $\dot R=(dR/d\eta
)(d\eta /dt)$ from Eqs. (\ref{R-tol-open}) and (\ref{t-tol-open}) obtaining 
\begin{equation}
\dot R(x,\eta )=\sqrt{f^2(x)-1}\frac{\sinh \eta }{\cosh \eta -1}=\sqrt{\frac{%
10}3\gamma a}x\frac{\sinh \eta }{\cosh \eta -1}  \label{dR/dt(eta)}
\end{equation}
where $a=|\xi ^{^{\prime \prime }}(0)|$. We therefore have 
\begin{equation}
\dot a(t)=\sqrt{\frac{10}3\gamma a}\frac{\sinh \eta }{\cosh \eta -1}
\label{da/dt}
\end{equation}
Replacing Eq. (\ref{da/dt}) in Eq. (\ref{Omega(t)}) we get 
\begin{equation}
\Omega =1-\tanh ^2\left( \frac \eta 2\right)  \label{Omega-abi}
\end{equation}
Evaluating $\eta $ from Eq. (\ref{t-tol-open}) with $t=t_{today}$ we get $%
\Omega _{today}\simeq 0.49$ at $x=0$.

\newpage\ 

{\bf FIGURE CAPTIONS}

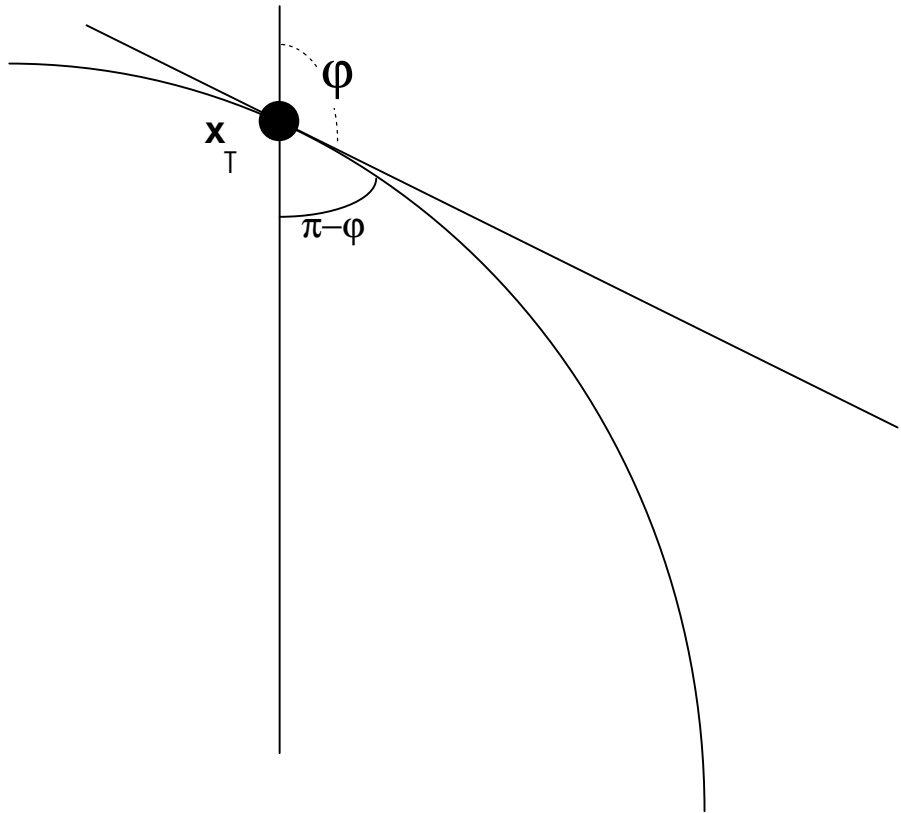
{\bf Fig. 1}: Angle subtended by the null geodesics that arrive at the
observer's position $x_T$.

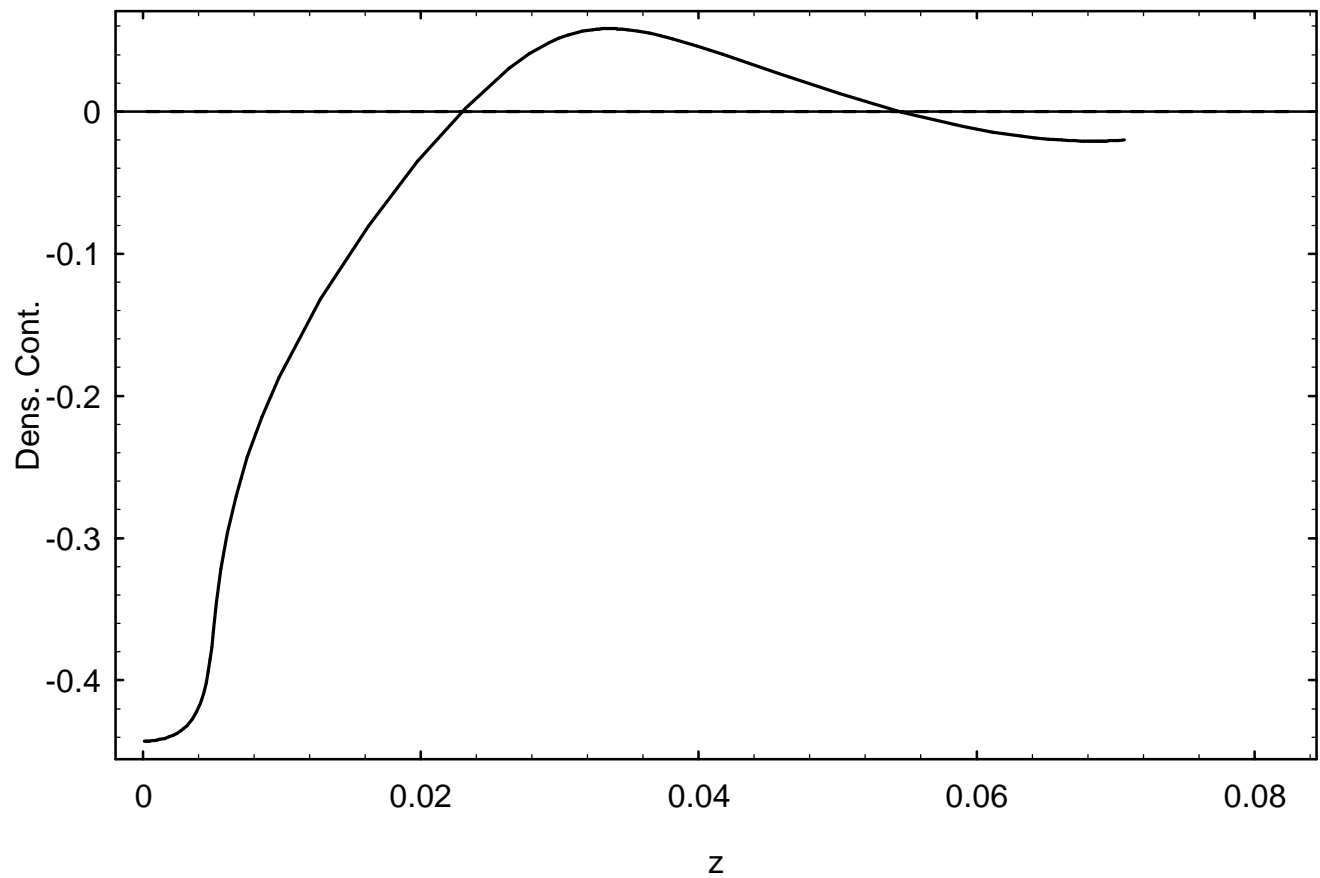
{\bf Fig. 2:} Density contrast $\Delta \delta /\bar \delta $ vs redshift $z$%
, to second order in the perturbative development. The profile corresponds
to a non compensated void.

{\bf Fig. 3:} $\ln N$ vs $\ln z$, to second order in the perturbative
development. Dashed line corresponds to flat FRW Universe, full line to
Tolman Universe

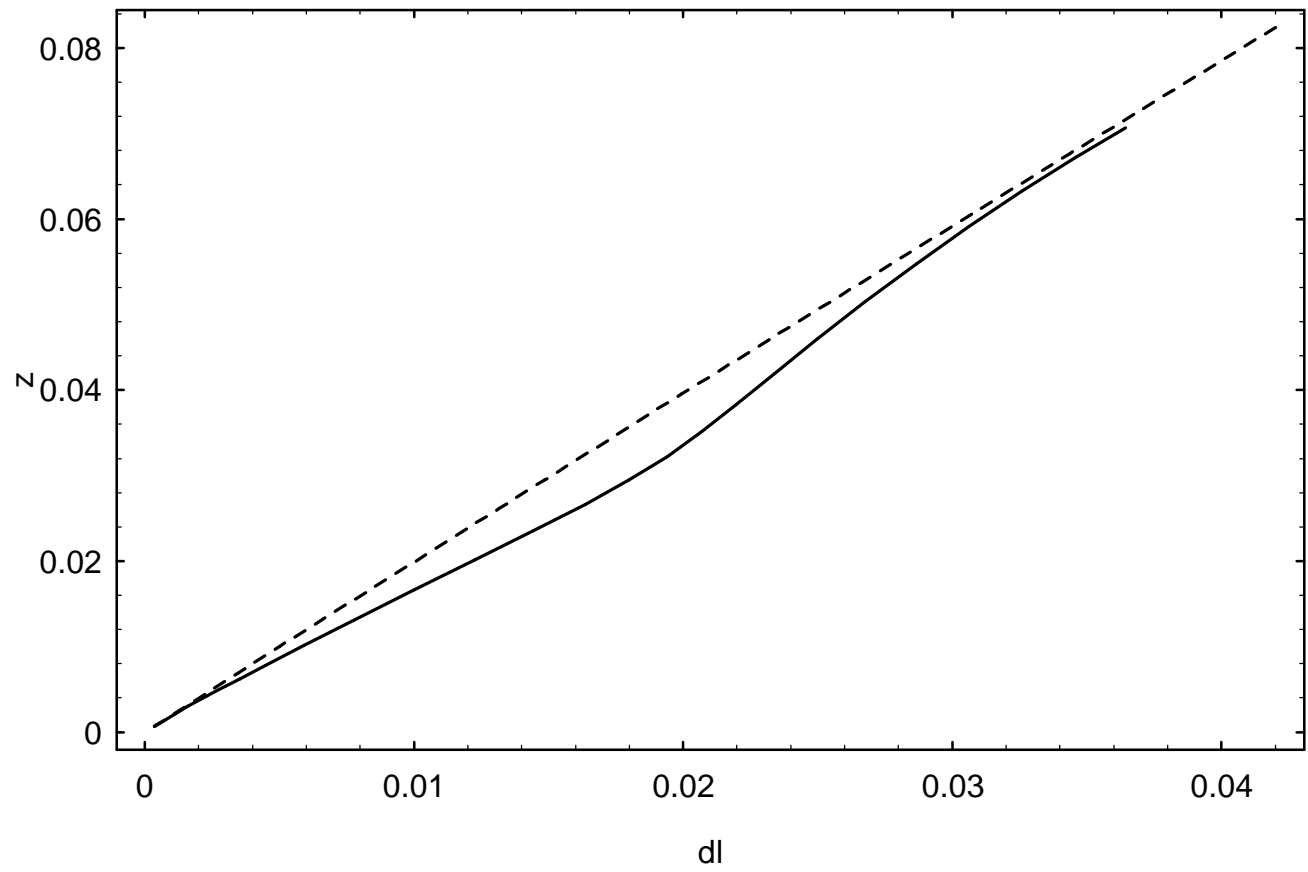
{\bf Fig. 4: }Redshift $z$ vs luminosity distance $dl$, to second order in
the perturbative development. Due to the small redshifts considered, the
slope of the curve is a measure of the Hubble constant. Dashed line
corresponds to flat FRW Universe, full line to Tolman Universe.


\begin{references}
\bibitem{Zeldovich}  Ya. B. Zel'dovich,Astron.Astrophys {\bf 5}, 84 (1970).

\bibitem{Roman}  Scoccimarro R., Frieman J., Fermilab-Pub-95/294-A (1995).

\bibitem{Shandarin}  Shandarin S. F., Zeldovich Ya. B., Rev. Mod. Phys {\bf %
61}, 185 (1989).

\bibitem{Hobill}  D. Hobill, A. Burd and A. Coley edts., {\it Deterministic
Chaos in General Relativity}, NATO ASI Series, Plenum Press, New York (1994).

\bibitem{MacCallum}  Kramer D., Stephani H., Herlt E., MacCallum M. A. H.; 
{\it Exact Solutions of Einstein Field Equations}, Cambridge University
Press, Cambridge (1980).

\bibitem{Krasinski}  A. Krasinski, {\it Physics in an Inhomogeneous Universe}%
, University of Cape Town preprint (1993).

\bibitem{voids}  Blumenthal G. R., Nicolaci Da Costa L., Goldwirth D. S.,
Lecar M., Piran T., Ap. J.{\bf \ 388}, 234 (1992).

\bibitem{Paczynski}  Paczy\'nski B, Piran T.; Ap. J. {\bf 364}, 341, (1990).

\bibitem{Panek}  Panek M., Ap. J. {\bf 388}, 225, (1992).

\bibitem{Ribeiro}  Ribeiro, M. B., ApJ{\bf . 388}, 1 (1992); ApJ{\bf . 395},
29 (1992); ApJ{\bf . 415}, 469 (1993).

\bibitem{Moffat-Tatarski}  Moffat J. W. \& Tatarski D. C., Phys. Rev. D {\bf %
45}, 3512 (1992).

\bibitem{Fullana}  Fullana M. J., Arnau J. V. S\'aez D.; astro-ph/9601154
(1996).

\bibitem{langlois}  Langlois D., Piran T., Phys. Rev. D {\bf 53}, 2908
(1996).

\bibitem{Campos}  Campos A., eprint astro-ph/9510051 (1995).

\bibitem{meszaros}  Meszaros P., Ast. Astrophys., {\bf 37}, 225 (1974).

\bibitem{BranFelMu92}  Brandenberger R. H., Feldman H. A. \& Mukhanov V. F.,
Phys. Rep. {\bf 215}, 203 (1992).

\bibitem{Kolb-Turner}  Kolb E. W., Turner M. S., {\it The Early Universe}
Addison-Wesley, New York (1990).

\bibitem{Borner}  B\"orner G., {\it The Early Universe}, Springer Verlag,
Berlin (1992).

\bibitem{Peebles93}  Peebles, P.J.E., {\it Principles of Physical Cosmology}
(New Jersey, Princeton Univ. Press) (1993).

\bibitem{Smoot}  Smoot G. F., ApJ. {\bf 396}, L1 (1992).

\bibitem{Land-Lifsh}  Landau L., Lifshitz E.M., {\it The Classical Theory of
Fields}, 4th edition, Oxford, Pergamon (1975).

\bibitem{Hell-Lake1}  Hellaby C., Lake K., Ap. J.{\bf 282}, 1 (1984).

\bibitem{Ellis}  Ellis G. F. R., in Relativistic Cosmology, Proc.
International School of Physics ``Enrico Fermi'', General Relativity and
Cosmology, R.K. Sachs edt., New York, Academic (1971).

\bibitem{Weinberg72}  Weinberg S., {\it Gravitation and Cosmology}, J. Wiley
\& Sons New York (1972).

\bibitem{Gravitation}  Misner C. W., Thorne K. S. \& Wheeler J. A., {\it %
Gravitation} (San Francisco, Freeman) (1973).

\bibitem{Steinhardt}  Steinhardt P.J.; astro-ph/9502024 (1995).

\bibitem{Gradshteyn}  Gradshteyn I. S. \& Ryzhik I. M., Table of Integrals,
Series and Products (New York, Academic) (1980).

\bibitem{partridge}  Partridge R. B., Class. Quantum Grav. {\bf 11}, A153
(1994).

\bibitem{deLapparent}  de Lapparent V., Geller M. J., Huchra J. P., ApJ {\bf %
302}, L1 (1986).
\end{references}
\end{document}